\documentclass[preprint,12pt]{elsarticle}

\usepackage{graphicx}
\usepackage{multicol}
\usepackage{amsmath,amssymb,amsfonts, mathtools}
\usepackage{mathrsfs}
\usepackage{float}
\usepackage{amsthm}  
\usepackage{rotating}
\usepackage[title, toc]{appendix}
\usepackage[numbers]{natbib}
\usepackage[a4paper,margin=1in]{geometry}
\usepackage{tikz, siunitx, newunicodechar}
\usepackage{booktabs,multirow}
\usepackage{tcolorbox}
\tcbuselibrary{skins,breakable}
\usepackage{xcolor}
\usepackage{lipsum}
\usepackage{longtable}
\usepackage{bm,bbm}%
\usepackage{hyperref}
\usepackage{rotating}
\usepackage{makecell}
\allowdisplaybreaks[4]

\theoremstyle{oupplain}
\newtheorem{theorem}{Theorem}[section]
\newtheorem{lemma}[theorem]{Lemma}
\newtheorem{proposition}[theorem]{Proposition}
\newtheorem{corollary}[theorem]{Corollary}
\theoremstyle{oupdefinition}
\newtheorem{definition}{Definition}[section]
\theoremstyle{oupremark}
\newtheorem{remark}[theorem]{Remark}
\newtheorem{example}[theorem]{Example}
\newtheorem{construction}[theorem]{Construction}
\theoremstyle{oupproof}

\journal{Finite Field and Their Applications}

\begin{document}

\begin{frontmatter}



\title{Function-Correcting Codes for $b-$Symbol Read Channels}


\author[1]{Anamika Singh}
\ead{anamikabhu2103@gmail.com} 

\affiliation[1]{organization={Department of Mathematics and Computing, Indian Institute of Technology (ISM) Dhanbad},
            addressline={Dhanbad, India}}
            
\author[2]{Abhay Kumar Singh \corref{cor1}}
\ead{itbhu81@gmail.com}
\affiliation[2]{organization = {Department of Mathematics and Computing, Indian Institute of Technology (ISM) Dhanbad},
             addressline ={Dhanbad, India} }
\cortext[cor1]{Corresponding Author}

\author[3]{Eitan Yaakobi}
\ead{yaakobi@cs.technion.ac.in}
\affiliation[3]{organization={Department of Computer Science, Technion- Israel Institute of Technology},
addressline = {Haifa, Israel} }
\begin{abstract}
Function correcting codes are an innovative class of codes that is designed to protect a function evaluation of the data against errors or corruptions. Due to its usefulness in machine learning applications and archival data storage, where preserving the integrity of computations is crucial, Lenz et al. \cite{lenz2023function} recently introduced function-correcting codes for binary symmetric channels to safeguard function evaluation against errors. Xia et al. \cite{xia2024function} expanded this concept to symbol-pair read channels over binary fields. The current paper further advances the theory by developing function-correcting codes for b-symbol read channels over finite fields. We introduce the idea of \textit{irregular $b-$symbol distance codes} and establish bounds on their performance over finite fields. This concept helps in understanding the behaviour of function-correcting codes in more complex settings. We use the connection between function-correcting $b-$symbol codes and irregular $b-$symbol distance codes to obtain bounds on its optimal redundancy. We also present a graphical representation of the problem of constructing function-correcting $b-$symbol codes. Furthermore, we apply these general results to specific classes of functions and compare the redundancy of function-correcting $b$-symbol codes with classical $b$-symbol codes. Our findings demonstrate that function-correcting $b$-symbol codes achieve lower redundancy while maintaining reliability.
\end{abstract}


\end{frontmatter}



\section{Introduction}
\label{sec:introduction}

    Classical error-correcting codes are typically designed to recover the entirety of a message transmitted over a noisy channel. However, in certain scenarios, the focus shifts to protecting only the function evaluation of the message rather than the message itself. This need inspired the development of function-correcting codes that would help to minimize redundancy, offering a more efficient alternative to the classical error-correcting codes that have been conventionally utilized. This new paradigm is based on the assumption that the sender knows the function and can use it to encode the message so that the recipient can reliably recover the value of the function. The key idea behind designing such codes is that the receiver need not distinguish between codewords from information vectors that evaluate to the same function value. When the function is injective, the function-correcting codes behave similarly to the classical error-correcting codes because each distinct input maps to a distinct output. In this case, function-correcting codes can be considered a generalization of classical error-correcting codes, where the primary focus is on preserving the function's evaluation rather than the original message. This new concept is suggested by Lenz et al. \cite{lenz2023function} for binary symmetric channels. In this paper, the relationship between irregular-distance codes and function-correcting codes is explored. This relationship is significant because it helped establish both lower and upper bounds on the redundancy of function-correcting codes. This is crucial because understanding these bounds can provide insights into the efficiency and effectiveness of these codes in different scenarios. Later, Xia et al. studied these codes for symbol-pair read channels over binary fields \cite{xia2024function}. Premlal and Ranjan \cite{premlal2024functioncorrectingcodes} focused on using function-correcting codes to correct linear functions.

    Building on this foundation, we generalize the coding framework of function-correcting codes from the symbol-pair read channels to the more expansive $b-$symbol read channels. Furthermore, we have broadened the domain of the functions from the binary fields to general finite fields, providing a more versatile and comprehensive coding approach.

    In this paper, we propose a graphical approach to the problem of constructing function-correcting $b-$symbol codes. An independent set of $q^k$ elements has been shown to meet the condition of function-correcting $b-$symbol codes. We use this approach to come up with a function-correcting $b-$symbol code for the $b-$ weight function. We establish a relationship between the shortest length of irregular $b-$symbol codes and the optimal redundancy of function-correcting $b-$symbol codes in the case of generic functions. We find bounds on the shortest length of irregular $b-$symbol distance codes and irregular distance codes for a larger finite field, as they in turn help to establish bounds on the optimal redundancy of function-correcting $b-$symbols. Then these results are applied to specific functions. Finally, we compare the redundancy of the function-correcting $b-$symbol codes for these functions with schemes that use classical $b-$symbol codes for $b = 3$.
    
    Function-correcting $b-$symbol codes offer a significant advantage over function-correcting codes (FCC) and function-correcting symbol-pair codes (FCSPC) in practical applications, particularly because modern storage devices increasingly rely on $b-$symbol read channels. These channels facilitate more efficient data storage and retrieval by enabling the processing of larger data units in a single operation.
    
    By designing codes specifically for $b-$symbol read channels, it becomes possible to optimize storage density while effectively managing error rates. In contrast, symbol-pair read channels are limited in their adaptability as they focus on correcting errors in smaller, fixed-size data units. These limitations underscore the importance of extending the theoretical framework of function-correcting codes for $b-$symbol read channels to address the needs of contemporary data storage systems more effectively.

    \begin{figure}
         \centering
         \includegraphics[width=0.8\linewidth]{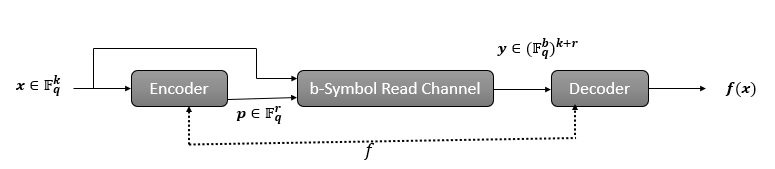}
         \caption{Setting for function correction in $b-$symbol read channel.}
         \label{fig:1}
     \end{figure} 
     

\section{Prelimnaries}
    Throughout this paper, let $q$ be a prime power and $\mathbb{F}_q$ be the finite field containing $q$ elements. We use $\mathbb{N}$ and $\mathbb{N}_0$ to denote the set of natural numbers and non-negative integers, respectively. For any matrix $\boldsymbol{B}$, we use $[\boldsymbol{B}]_{ij}$ to indicate the $(i,j)th$ entry of $\boldsymbol{B}$.
    
    In this section, we discuss some foundational concepts related to $b-$symbol read channels ($2 < b < k)$ and irregular-distance codes \cite{lenz2023function}, both of which are important for developing function-correcting codes for these channels. 


\subsection{$b-$Symbol Codes} 
    
        High-density data storage technologies do not allow individual symbol-read operations off the channel. To address this limitation, Cassuto and Blaum proposed codes for symbol-pair read channels \cite{5983443}, where two overlapping symbols are read simultaneously instead of just one. After the seminal work by Cassuto and Blaum, the field has seen significant advancements that have enriched the research area of symbol-pair codes. See \cite{chee2012maximum, dinh2017symbol,chen2017constacyclic, TANG2023102203}. Yaakobi et al. expanded the symbol-pair framework to the $b-$symbol model \cite{7393834}, where groups of symbols are read together$(b > 2)$. 
        To understand the implementation of function-correcting codes for the $b-$symbol read channels where $2 < b < k $, we review the $b-$symbol read channels model and their basic properties \cite{yang2016construction, mostafanasab2016b,ding2018maximum, TANG2023102203}.
        
         \begin{definition}Let $\boldsymbol{x} = (x_0, x_1, \ldots, x_{k-1})$  be a vector in $\mathbb{F}_q^k$. Then $b-$symbol read vector $\pi_b(\boldsymbol{x})$ of $\boldsymbol{x}$ is given by:
        \[
            \pi_b(\boldsymbol{x}) = [ (x_0, x_1, \ldots, x_{b-1}), (x_1, x_{2}, \ldots, x_{b}), \ldots , (x_{k-1}, x_0, \ldots, x_{b-2}) ].
        \]
        \end{definition}

        After defining the $b$-symbol read vector, we introduce the concepts of $b$-symbol distance and $b$-symbol read codes.
        
        \begin{definition}
            For any vectors $\boldsymbol{x}, \boldsymbol{y} \in \mathbb{F}_q^k$, the $b-$symbol distance between $\boldsymbol{x}$ and $\boldsymbol{y}$ is defined as
            \[
                d_b(\boldsymbol{x},\boldsymbol{y}) = d_H(\pi_b(\boldsymbol{x}), \pi_b(\boldsymbol{y})).
            \]
        \end{definition}
        
        \begin{definition} For a code $\mathcal{C}$ over $\mathbb{F}_q$, the $b-$symbol read code of $\mathcal{C}$ is a code over  $\mathbb{F}_q^b$ and is defined as follows:
        \[
            \pi_b(\mathcal{C}) = \{\pi_b(\boldsymbol{c}) : \boldsymbol{c} \in \mathcal{C}\}
        \]
               The minimum $b$-distance of the code, $d_b(\mathcal{C})$, can be determined using the formula  
        \[
            d_b(\mathcal{C}) = d_H (\pi_b(\mathcal{C})),
        \]
        where $d_H (\pi_b(\mathcal{C}))$ is the minimum Hamming distance of the code $\pi_b(\mathcal{C})$. Moreover, $\mathcal{C}$ can correct any $t$ $b$-symbol errors \cite{7393834} iff  
        \[
            d_b(\mathcal{C}) \geq 2t + 1. 
        \]
         \end{definition}
        
         The following lemma discusses the relationship between the Hamming distance and the $b-$symbol distance, between two vectors.
         
         \begin{lemma} \label{hamming_b_rel} Let $ \boldsymbol{x}$ and $ \boldsymbol{y} $ be two vectors in $\mathbb{F}_q^k$ with Hamming distance $ d_H(\boldsymbol{x}, \boldsymbol{y}) $. The $ b-$symbol distance between them, $ d_b(\boldsymbol{x}, \boldsymbol{y}) $, satisfies the following conditions:
        \begin{itemize}
            \item If $ 0 < d_H(\boldsymbol{x}, \boldsymbol{y}) \leq k - (b - 1) $, then
            \[
                d_H(\boldsymbol{x},\boldsymbol{y}) + b - 1 \leq d_b(\boldsymbol{x},\boldsymbol{y}) \leq b \cdot d_H(\boldsymbol{x},\boldsymbol{y}).
            \]
            \item If $ d_H(\boldsymbol{x}, \boldsymbol{y}) > k - b + 1 $, then
            \[
                d_b(\boldsymbol{x},\boldsymbol{y}) = k.
            \]
            \item If $ d_H(\boldsymbol{x}, \boldsymbol{y}) = 0 $, then
            \[
                d_b(\boldsymbol{x},\boldsymbol{y}) = 0.
            \]
        \end{itemize}
        \end{lemma}

         Next, we give a relation that connects $(b+1)-$symbol distance and $b-$symbol distance between two vectors in $\mathbb{F}_q^k$

        \begin{proposition}\label{b+1_and_b}
        For any two vectors $\bm{x} = (x_0, \ldots, x_{k-1})$ and $\bm{y} = (y_0, \ldots, y_{k-1})$ in $\mathbb{F}_q^k$ and $0 < d_b(\bm{x,\bm{y}}) <  k$ , we have
        \[
                d_{b+1}(\boldsymbol{x},\boldsymbol{y})  \geq d_b(\boldsymbol{x},\boldsymbol{y}) + 1.
        \]
        \end{proposition}

        \begin{proof}
        For any $0 \leq i \leq k-1$, if $ (x_i, \ldots, x_{i+b-1}) \neq (y_i, \ldots, y_{i+b-1}) $, then $ (x_i, \ldots, x_{i+b-1},x_{i+b}) \neq (y_i, \ldots, y_{i+b-1}, y_{i+b})$. Hence, $ d_{b+1}(\boldsymbol{x},\boldsymbol{y})  \geq d_b(\boldsymbol{x},\boldsymbol{y}) $. Also, since $d_b(\boldsymbol{x},\boldsymbol{y}) <  k$ there must exist an index $j$ such that $(x_j, \ldots, x_{j+b-1}) = (y_j, \ldots, y_{j+b-1}) $ and $x_{j+b} \neq y_{j+b}$. It follows that $(x_j, \ldots, x_{j+b-1}, x_{j+b}) \neq (y_j, \ldots, y_{j+b-1}, y_{j+b}) $, therefore $d_{b+1}(\boldsymbol{x},\boldsymbol{y})  \geq d_b(\boldsymbol{x},\boldsymbol{y}) + 1.$
        \end{proof}

        For any two vectors \(\boldsymbol{x} = (x_1, \ldots, x_k)\) and \(\boldsymbol{y} = (y_1, \ldots, y_k)\) in \(\mathbb{F}_q^k\) satisfying \(0 < d_H(\boldsymbol{x},\boldsymbol{y}) < k - b + 2\), let  $I_H = \{ i : x_i \neq y_i \}$ denote the set of indices where \(\boldsymbol{x}\) and \(\boldsymbol{y}\) differ. Consider a minimal partition of \(I_H\) into subsets of cyclically consecutive indices (where the indices wrap around the modulo \(k\)), denoted as $I_H = \bigcup_l B_l.$\\  
        Now, take two consecutive subsets in this partition, say $B_i = \{p, p+1, \ldots, q\} \quad \text{and} \quad B_{i+1} = \{r, r+1, \ldots, s\},$ where all indices between \(q\) and \(r\) satisfy \(x_i = y_i\). Observing the structure of the \(b\)-symbol read vector, we note that the coordinate \(x_r\) appears in exactly \(b\) different \(b\)-coordinate vectors, namely:  
        \[
        (x_{r-b+1}, \ldots, x_r), \quad (x_{r-b+2}, \ldots, x_r, x_{r+1}), \quad \ldots, \quad (x_r, \ldots, x_{r+b-1}).
        \]
        As a result, the subset \(B_{i+1}\) contributes at most \(|B_{i+1}| + (b - 1)\) to the \(b\)-symbol distance between \(\boldsymbol{x}\) and \(\boldsymbol{y}\). However, if the number of indices between \(q\) and \(r\) is strictly less than \(b-1\), then \(B_{i+1}\) contributes only $|B_{i+1}| + (r - q - 1)$ to the \(b\)-symbol distance, since the remaining contribution has already been accounted for by \(B_i\).\\
        Using this observation, we can express the \(b\)-symbol distance in terms of the Hamming distance, as stated in the following proposition.
        
        \begin{proposition} For any two vectors $\boldsymbol{x}$ and $\boldsymbol{y}$ in $\mathbb{F}_q^k$ whose Hamming distance satisfies the inequality $0 < d_H(\boldsymbol{x}, \boldsymbol{y}) < k - b + 2$, define the subset of $[1, k]$ consisting of all the positions where $\boldsymbol{x}$ and $\boldsymbol{y}$ are equal as $I = \{ i : x_i = y_i\}$. Let $ I = \bigcup A_i$ be a minimal partition of the set $I$ into subsets of cyclically consecutive indices. If $A = \{ i : |A_i| \geq b-1\}$ then,
        \[
            d_b(\boldsymbol{x}, \boldsymbol{y}) = d_H(\boldsymbol{x}, \boldsymbol{y}) + (b-1)L + e.
        \]
where $L = |A|$ and $e = \sum_{i\in I-A}|A_i|$
        \end{proposition}

        \begin{example}
            Let $k = 11, b = 3$, $x = (1,0,1,0,1,1,0,0,0,0,0)$ and $y = (0,0,0,0,0,0,0,0,0,0,0)$ then $L = 1 $, $e = 2$ and the Hamming distance between them is equal to  4  . The $3-$symbol read vector of $\boldsymbol{x}$ and $\boldsymbol{y}$ is given as follows:
            \begin{align*}
                    \pi_b(\boldsymbol{x}) = [ (1,0,1), (0,1,0), (1,0,1), (0,1,1), (1,1,0), (1,0,0), \\(0,0,0), (0,0,0), (0,0,0), (0,0,1),(0,1,0)]
            \end{align*} 
            \begin{align*}
                        \pi_b(\boldsymbol{y}) = [(0,0,0), (0,0,0), (0,0,0), (0,0,0), (0,0,0), (0,0,0), \\(0,0,0), (0,0,0), (0,0,0), (0,0,0), (0,0,0)]
            \end{align*} 
            From the above sequence of $3-$symbol coordinate vectors of $\boldsymbol{x}$ and $\boldsymbol{y}$ we can clearly see that the $b-$symbol distance between them is $8.$ By the formula given in Proposition 2.2, we get 
            \[ d_b(\boldsymbol{x}, \boldsymbol{y}) = 4 + 1 \times 2 + 2 = 8.\]
        \end{example}
        
        $b-$symbol sphere $\mathcal{S}^b_k(t)$ of radius $t$ is a structure representing a set of sequences over $\mathbb{F}_q^k$ which are at $b-$symbol distance $t$ from a central sequence.        
        For $k > l \geq L_j$, Proposition 2.3 gives an enumerative formula to calculate the size of $b-$symbol sphere of radius $t$ for $b=3$, with the help of the expression $D_j(k, l, L_j) = \frac{k}{L_j}\binom{l-1}{L_j-1}\binom{k-l-L_j-1}{L_j-j-1}\binom{L_j}{j}$. Here, $D_j(n, l, L_j) $ gives the count of the number of sequences that have length $k$ with Hamming distance $l$ and minimal partition of the set $I$ equal to $L_j$, where $L_j$ is the number of subsets in a partition with exactly $j$  subsets of size $1$ and rest that have size greater than or equal to $2( b-1 = 2$  for $b = 3)$. The proof of the expression of $D_j(k, l, L_j)$ is given in \ref{appendix}. 
        
        \begin{proposition}\label{b-symbol_ball} For $b = 3$, any $\boldsymbol{x} \in \mathbb{F}_q^k$ and $0 < i < k$
        \[
            |\mathcal{S}^3_k(i)| =\sum_{j = 0}^{\lfloor \frac{k}{2} \rfloor} \sum_{l = \lceil \frac{i+j}{3} \rceil}^{i-1} D_j(k, l, \frac{i-l+j}{2})(q-1)^l.
        \]
        and
        \[
            |\mathcal{B}^3_k(t)| = 1 + \sum^t_{i=1}|\mathcal{S}^3_k(i)|.
        \]
where $\mathcal{B}^3_k(t)$ is the $3-$symbol ball of radius $t$ and represents all sequences of length $k$ that are at $b-$symbol distance $t$ or less from a central sequence.
        \end{proposition}
        
        \begin{lemma}\label{sphere_packing}\cite{song2018sphere} According to the $b$-symbol sphere packing bound for any $q$-ary $b$-symbol code $\mathcal{C} \subseteq \mathbb{F}_q^k$ that can correct at most $t$-errors and has the code size M satisfies
        \[
            M|\mathcal{B}^b_k(t)| \leq q^k.
        \]
        \end{lemma}


   \subsection{Irregular-Distance Codes}
        \begin{definition} \cite{premlal2024function} A set of codewords $\mathcal{P} =\{ \boldsymbol{p_1}, \ldots, \boldsymbol{p_M} \}$ is said to be a $\boldsymbol{B}-$Irregular$-$distance code ( $\boldsymbol{B}$-code ) for some matrix $\boldsymbol{B}$ $\in \mathbb{N}_0^{M \times M}$ if there exists an ordering of the codewords in $\mathcal{P}$ such that $d_b(\boldsymbol{p_i},\boldsymbol{p_j}) \geq$ $\boldsymbol{[B]_{ij}}$ for all $i,j \in \{1, 2, \ldots , M\}.$\\
        \end{definition}
        $N_H(\boldsymbol{B})$  denotes the smallest positive integer $r$  for which there exists a $\boldsymbol{B}-$code of length $r.$  In particular, when all non-diagonal entries are equal to some non-negative integer D, we use the notation $N_H(M, D) $ in place of $N_H(\boldsymbol{B}).$

\section{Function-Correcting Codes for $b-$Symbol Read Channel} 

    Let $\boldsymbol{x} = ({x_0}, {x_2},.., {x_{k-1}})$ be an information vector of length $k$ over ${\mathbb{F}_q}$. Consider a function $f: {\mathbb{F}^k_q} \rightarrow Im(f)$, where $Im(f) = \{f(\boldsymbol{x}): \boldsymbol{x} \in {\mathbb{F}^k_q} \}$ denotes the image of $f$. Let $E = |Im(f)|$ be the cardinality of the image of $f$.
    The information vector $\boldsymbol{x}$ is encoded using a systematic encoding function 
    \begin{equation*}
        Enc: {\mathbb{F}^k_q} \rightarrow\mathbb{F}_q^{k+r}, \quad Enc(\boldsymbol{x}) = (\boldsymbol{x},p(\boldsymbol{x}))
    \end{equation*}

    where $p(\boldsymbol{x}) \in\mathbb{F}_q^r$ is the redundancy vector, and $r$ represents the length of redundancy added to $\boldsymbol{x}$. After encoding, the codeword is sent over an erroneous $b-$symbol read channel. Assuming that at most $t$ errors can occur in codewords during transmission where $t > \lfloor \frac{b-1}{2} \rfloor$, we define function-correcting $b$-symbol codes for the function $f$ as follows.
    \begin{definition}An encoding function, $ Enc: {\mathbb{F}^k_q} \rightarrow {\mathbb{F}_q^{k+r}},  Enc(\boldsymbol{x}) = (\boldsymbol{x},p(\boldsymbol{x}))$ defines a \textbf{function correcting $b$-symbol} code for the function $f: \mathbb{F}^k_q \rightarrow Im(f)$  if 
    \begin{equation*}  
        d_b(Enc(\boldsymbol{x_1}), Enc(\boldsymbol{x_2})) \geq 2t+1 \hspace{3mm}\forall \quad\boldsymbol{x_1}, \boldsymbol{x_2} \in {\mathbb{F}^k_q} \text{ with  } f(\boldsymbol{x_1}) \neq f(\boldsymbol{x_2})  
    \end{equation*}
    \end{definition}
     The main goal or interest is to achieve function value correction through function-correcting $b-$symbol code with as small redundancy as possible. Keeping this in mind, we define optimal redundancy.
     
    \begin{definition} The smallest positive integer $r$ for which a function-correcting $b$-symbol code exists with an encoding function $ Enc: {\mathbb{F}^k_q} \rightarrow {\mathbb{F}_q^{k+r}}$ for the function $f$ is called the \textbf{optimal redundancy} of $f$.  It is denoted as $r_b^f(k,t)$.
    \end{definition}


    \subsection{Equivalence of Function-Correcting $b-$Symbol Codes and Independent Set of Function-Dependent Graph}
    We define function-dependent graph whose independent sets, when of appropriate size, can be used to construct a function-correcting \( b \)-symbol code.

    \begin{definition}
        A function-dependent graph $G_f^b(k, t, r)$ is defined to be the graph whose vertex set $V$ is given by $V = \mathbb{F}_q^k \times \mathbb{F}_q^r$, such that any node in the graph has the form $\boldsymbol{v} = (\boldsymbol{x}, \boldsymbol{p}) \in \mathbb{F}_q^k \times \mathbb{F}_q^r.$  An edge will connect two nodes in $G_f^b(k, r, t)$,  $\boldsymbol{v_1} = (\boldsymbol{x_1}, \boldsymbol{p_1})$ and  $\boldsymbol{v_2} = (\boldsymbol{x_2}, \boldsymbol{p_2})$  iff
        \begin{itemize}
            \item $\boldsymbol{x_1} = \boldsymbol{x_2}$, or
            \item $f(\boldsymbol{x_1}) \neq f(\boldsymbol{x_2})$ and $d_b(\boldsymbol{v_1}, \boldsymbol{v_2}) <  2t + 1.$
        \end{itemize}
        \end{definition}

\begin{figure}[t]
     \centering
     \includegraphics[width=0.75\linewidth]{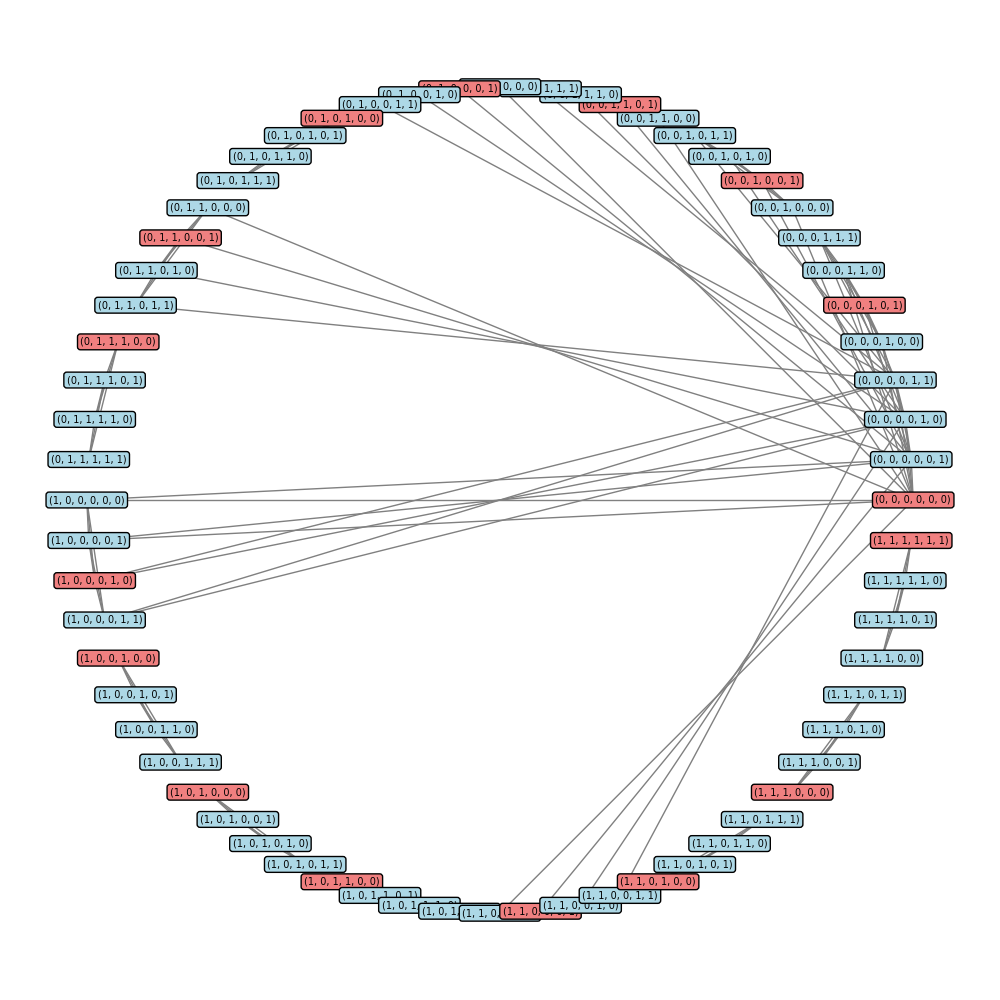}
     \caption{Graph $G_f^b(k,t,r)$ for $ b = 3, k = 4, t = 2, r = 2$ and function $f((x_1,x_2,x_3,x_4)) = x_1 \vee x_2 \vee x_3 \vee x_4$ over binary field.  The nodes in red boxes that are 16 in size form a function-correcting $b-$symbol code for function $f$}
     \label{Fig:2}
    \end{figure}

    An independent set of the graph $G_f^b(k, r, t)$ can be used to construct a function-correcting $b-$symbol code, provided that its size is $q^k$. This is because the first condition for edge formation ensures that no two identical information vectors are assigned different redundancy vectors, while the second condition guarantees that any two information vectors yielding different function values have a $b-$symbol distance of at least $2t+1$ between their corresponding codewords.\\
    We use the notation $\gamma_f^b(k,t)$ to denote the smallest integer $r$ such that there exists an independent set of size $q^k$ in $G_f^b(k,t,r).$ We can easily see that $r_b^f(k,t) = \gamma_f^b(k,t)$, for any generic function $f$ defined over $\mathbb{F}_q^k$ as the independent set of the graph $G_f^b(k, t, r)$ where $r = \gamma_f^b(k,t)$ satisfies all the conditions to be a function-correcting $b-$symbol code and vice versa.  Figure \ref{Fig:2} shows the graph $G_f^b(k, t, r)$ for $b = 3$ and the corresponding independent set that serves as a function-correcting $b-$symbol code for the function $f: \mathbb{F}_2^4 \rightarrow  \mathbb{F}_2$ defined by $f((x_1,x_2,x_3,x_4)) = x_1 \vee x_2 \vee x_3 \vee x_4.$

 \subsection{ Relation between Irregular $b-$Symbol Distance Codes and Function-Correcting $b-$Symbol Codes}

    To explore optimal redundancy, we establish the connection between function-correcting $b-$symbol codes and irregular $b-$symbol distance codes, analogous to the approach in \cite{lenz2023function} and \cite{xia2024function}.\\
    Before defining the irregular $ b-$symbol distance codes, we introduce two types of irregular $b-$symbol distance matrices associated with a function $f$.
    
     \begin{definition}: Consider $M$ vectors $\boldsymbol{x_0}, . . . , \boldsymbol{x_{M-1}} \in {\mathbb{F}^k_q}$. Then,  $\boldsymbol{B_f^{(1)}}(t, \boldsymbol{x_0}, . . . , \boldsymbol{x_{M-1}})$ and
    $\boldsymbol{B_f^{(2)}}(t, \boldsymbol{x}_0, . . . , \boldsymbol{x}_{M-1})$  are $M \times M$  $b-$symbol distance matrices with entries
 
    \begin{equation*}
        [\boldsymbol{B_f^{(1)}}(t, \boldsymbol{x}_0, . . . , \boldsymbol{x}_{M-1})]_{ij}=   
        \begin{cases}
           [2t - b + 2 - d_b(\boldsymbol{x}_i, \boldsymbol{x}_j)]^+ , & if \hspace{1mm} f (\boldsymbol{x}_i) \neq f(\boldsymbol{x}_j),\\0, & otherwise.
        \end{cases}
    \end{equation*}
    
    and 
    
    \begin{equation*}
        [\boldsymbol{B_f^{(2)}}(t, \boldsymbol{x}_0, . . . , \boldsymbol{x}_{M-1})]_{ij}=   
        \begin{cases}
           [2t + b - d_b(\boldsymbol{x}_i, \boldsymbol{x}_j)]^+ , & if \hspace{1mm} f (\boldsymbol{x}_i) \neq f(\boldsymbol{x}_j),\\0, & otherwise.
        \end{cases}
    \end{equation*}
    \end{definition}
    
    The \( [\cdot]^+ \) operator means that the value is nonnegative, that is, \( [x]^+ = \max\{x, 0\} \).
    When $M = q^k$, instead of writing $\boldsymbol{B_f^{(1)}}(t, \boldsymbol{x}_0, . . . , \boldsymbol{x}_{M-1})$ and $\boldsymbol{B_f^{(2)}}(t, \boldsymbol{x}_0, . . . , \boldsymbol{x}_{M-1})$, we use the shorthand  notations $\boldsymbol{B_f^{(1)}}(t)$ and $\boldsymbol{B_f^{(2)}}(t)$ respectively, for simplicity.\\
    We now formally define irregular $ b-$symbol distance codes as follows.
    
    \begin{definition}: A set of codewords $\mathcal{P} = \{ \boldsymbol{p_1}, \ldots, \boldsymbol{p_M} \}$ is said to be a $\boldsymbol{B}$-irregular $b-$symbol distance code ( $\boldsymbol{B_b}$-code ) for some matrix $\boldsymbol{B}$ $\in \mathbb{N}_0^{M \times M}$ if there exists an ordering of the codewords in $\mathcal{P}$ such that $d_b(\boldsymbol{p_i},\boldsymbol{p_j}) \geq$ $\boldsymbol{[B]_{ij}}$ for all $i,j \in \{1, 2, \ldots, M\}.$
    \end{definition}
    Let $N_b(\boldsymbol{B})$ denote the smallest positive integer $r$  for which there exists a $\boldsymbol{B_b}-$code of length $r$.  In particular, when all non-diagonal entries are equal to some non-negative integer D, we use the notation $N_b(M, D)$ in place of $N_b(\boldsymbol{B})$.\\  
    We now prove the lemma, which helps establish a relation between function-correcting $b-$symbol codes and irregular $ b-$distance codes of the matrices \(\boldsymbol{ B}_f^1(t) \) and \( \boldsymbol{B}_f^2(t) \).
    \begin{lemma}
    Let  $\boldsymbol{x} = (x^{(1)}, x^{(2)}) \in \mathbb{F}_q^{k+r}$  and $\boldsymbol{y} = (y^{(1)}, y^{(2)}) \in \mathbb{F}_q^{k+r}$, where
    \[
    x^{(1)} = (x_0, \ldots, x_{k-1}), \quad x^{(2)} = (x_{k}, \ldots, x_{k+r-1}),
    \]
    \[
    y^{(1)} = (y_0, \ldots, y_{k-1}), \quad y^{(2)} = (y_{k}, \ldots, y_{k+r-1}).
    \]
then, we have
    \[
    d_b(x^{(1)}, y^{(1)}) + d_b(x^{(2)}, y^{(2)}) - (b-1) \leq d_b(\boldsymbol{x},\boldsymbol{y}) \leq d_b(x^{(1)}, y^{(1)}) + d_b(x^{(2)}, y^{(2)}) + (b-1).
    \]
    \end{lemma}
    \begin{proof} The $b$-symbol distance between two vectors $\boldsymbol{x}$ and $\boldsymbol{y}$ is given by  
\begin{align*}
    d_b(\boldsymbol{x}, \boldsymbol{y}) &= d_H(\pi_b(\boldsymbol{x}), \pi_b(\boldsymbol{y}))\\
    &= d_H([({x}_0, \ldots, {x}_{b-1}), \ldots, ({x}_{k+r-1}, \ldots, {x}_{b-2})], \\
    &\quad [({y}_0, \ldots,{y}_{b-1}), \ldots, ({y}_{k+r-1}, \ldots, {y}_{b-2})]).
\end{align*}
Similarly, the $b$-symbol distances for $x^{(1)}$ and $y^{(1)}$, as well as $x^{(2)}$ and $y^{(2)}$, are given by  
\begin{align*}
    d_b(x^{(1)}, y^{(1)}) &= d_H(\pi_b(x^{(1)}), \pi_b(y^{(1)})) \\
    &= d_H([({x}_0, \ldots, {x}_{b-1}), \ldots, ({x}_{k-1}, \ldots, {x}_{b-2})], \\
    &\quad [({y}_0, \ldots, {y}_{b-1}), \ldots, ({y}_{k-1}, \ldots, {y}_{b-2})]).
\end{align*}  
\begin{align*}
    d_b(x^{(2)}, y^{(2)}) &= d_H(\pi_b(x^{(2)}), \pi_b(y^{(2)})) \\
    &= d_H([({x}_k, \ldots, {x}_{k+b-2}), \ldots, ({x}_{k+r-1}, \ldots, {x}_{k+b-2})], \\
    &\quad [({y}_k, \ldots, {y}_{k+b-2}), \ldots, ({y}_{k+r-1}, \ldots,{y}_{k+b-2})]).
\end{align*}

The discrepancy in $b$-symbol distances arises due to differences in the $b$-coordinate vectors at the boundary positions. Specifically, the $b$-coordinate vectors that appear in $\pi_b(\boldsymbol{x})$ and $\pi_b(\boldsymbol{y})$ but are absent in $(\pi_b(x^{(1)}), \pi_b(x^{(2)}))$ and $(\pi_b(y^{(1)}), \pi_b(y^{(2)}))$ are denoted as $\eta_x$ and $\eta_y$, respectively:
\begin{align*}
    \eta_x &= [({x}_{k-b+1}, \ldots, {x}_{k}), \ldots, ({x}_{k-1}, \ldots, {x}_{k+b-2}), ({x}_{k+r-b+1}, \ldots, {x}_{0}), \ldots, ({x}_{k+r-1}, \ldots, {x}_{b-2})], \\
    \eta_y &= [({y}_{k-b+1}, \ldots, {y}_{k}), \ldots, ({y}_{k-1}, \ldots, {y}_{k+b-2}), ({y}_{k+r-b+1}, \ldots, {y}_{0}), \ldots, ({y}_{k+r-1}, \ldots, {y}_{b-2})].
\end{align*}

Conversely, the $b$-coordinate vectors that appear in $(\pi_b(x^{(1)}), \pi_b(x^{(2)}))$ and $(\pi_b(y^{(1)}), \pi_b(y^{(2)}))$ but are absent in $\pi_b(\boldsymbol{x})$ and $\pi_b(\boldsymbol{y})$ are denoted as $\zeta_{\Bar{x}}$ and $\zeta_{\Bar{y}}$, respectively:
\begin{align*}
    \zeta_{\Bar{x}} &= [({x}_{k-b+1}, \ldots, {x}_{0}), \ldots, ({x}_{k-b+2}, \ldots, {x}_{1}), ({x}_{k-1}, \ldots, {x}_{b-2}), \ldots, ({x}_{k+r-1}, \ldots, {x}_{k+b-2})], \\
    \zeta_{\Bar{y}} &= [({y}_{k-b+1}, \ldots,{y}_{0}), \ldots, ({y}_{k-1}, \ldots, {y}_{b-2}), ({y}_{k+r-b+1}, \ldots, {y}_{0}), \ldots, ({y}_{k+r-1}, \ldots, {y}_{b-2})].
\end{align*}

From the above definitions, we obtain:
\begin{align*}
    d_b(\boldsymbol{x},\boldsymbol{y}) - d_b(x^{(1)}, y^{(1)}) - d_b(x^{(2)}, y^{(2)}) = d_H(\eta_x, \eta_y) - d_H(\zeta_{\Bar{x}}, \zeta_{\Bar{y}}).
\end{align*}
Since,
\begin{align*}
    0 \leq d_H(\eta_x, \eta_y) \leq 2(b-1), \quad 0 \leq d_H(\zeta_{\Bar{x}}, \zeta_{\Bar{y}}) \leq 2(b-1),
\end{align*}
we obtain:
\begin{align*}
   -2(b-1) \leq d_H(\eta_x, \eta_y) - d_H(\zeta_{\Bar{x}}, \zeta_{\Bar{y}}) \leq 2(b-1).
\end{align*}
Next, suppose that
\begin{align*}
    b \leq d_H(\eta_x, \eta_y) - d_H(\zeta_{\Bar{x}}, \zeta_{\Bar{y}}) \leq 2(b-1).
\end{align*}
This would imply
\begin{align*}
    0 \leq d_H(\zeta_{\Bar{x}}, \zeta_{\Bar{y}}) \leq b - 2.
\end{align*}
However, this condition requires that 
\begin{align*}
    d_H(\eta_x, \eta_y) = 0,
\end{align*}
which leads to a contradiction. Therefore, we conclude that
\begin{align*}
    0 \leq d_H(\eta_x, \eta_y) - d_H(\zeta_{\Bar{x}}, \zeta_{\Bar{y}}) \leq (b-1).
\end{align*}

Similarly, if
\begin{align*}
    -2(b-1) \leq d_H(\eta_x, \eta_y) - d_H(\zeta_{\Bar{x}}, \zeta_{\Bar{y}}) \leq -b,
\end{align*}
this condition is also not possible, leading to
\begin{align*}
    -(b-1) \leq d_H(\eta_x, \eta_y) - d_H(\zeta_{\Bar{x}}, \zeta_{\Bar{y}}) \leq 0.
\end{align*}

Combining both inequalities, we obtain
\begin{align*}
    -(b-1) \leq d_H(\eta_x, \eta_y) - d_H(\zeta_{\Bar{x}}, \zeta_{\Bar{y}}) \leq b-1.
\end{align*}
which further implies
\begin{align*}
    d_b(x^{(1)}, y^{(1)}) + d_b(x^{(2)}, y^{(2)}) - (b-1) \leq d_b(x, y) \leq d_b(x^{(1)}, y^{(1)}) + d_b(x^{(2)}, y^{(2)}) + (b-1).
\end{align*}\end{proof}
    
    Now that the above lemma has been proven, we can prove the following theorem which establishes the relation between function-correcting $b-$symbol code and irregular $b-$symbol distance codes of the matrices $\boldsymbol{B}_f^1(t)$ and $\boldsymbol{B}_f^2(t)$.
    
    \begin{theorem}\label{Theorem 3.2} For any function $f: \mathbb{F}^k_q \rightarrow Im(f)$ and $\{\boldsymbol{x_1}, \ldots, \boldsymbol{x_{q^k}}\} = \mathbb{F}^k_q$, we have 
    \[
        N_b(\boldsymbol{B}_f^{(1)}(t)) \leq r_b^f(k,t) \leq N_b(\boldsymbol{B}_f^{(2)}(t)).
    \]
    \end{theorem}
    \begin{proof} We establish the theorem by considering the following cases.\\
Case 1: Constant functions
If \( f \) is a constant function, then both \( N_b(\boldsymbol{B}_f^{(1)}(t)) \) and \( N_b(\boldsymbol{B}_f^{(2)}(t)) \) are equal to zero. Consequently, the desired condition holds trivially, establishing the theorem in this case. Specifically, we have:  
    \[
    N_b(\boldsymbol{B}_f^{(1)}(t)) = r_b^f(k,t) = N_b(\boldsymbol{B}_f^{(2)}(t)).
    \]
Case 2: Non-constant functions
To prove that \( N_b(\boldsymbol{B}_f^{(1)}(t)) \leq r_b^f(k,t) \) for non-constant functions \( f \), consider a function-correcting \( b \)-symbol code defined by an encoding function
\[
    \Phi: \mathbb{F}_q^k \to \mathbb{F}_q^{k+r}, \quad \boldsymbol{x_i} \mapsto (\boldsymbol{x_i}, \boldsymbol{p_i}),
 \]
where the redundancy \( r \) is optimal, i.e.,  \( r = r_b^f(k,t) \). Suppose, for contradiction, that \( N_b(\boldsymbol{B}_f^{(1)}(t)) > r_b^f(k,t) \). This implies the existence of distinct indices \( i,j \in \{1, \ldots, q^k\} \) such that \( f(\boldsymbol{x_i}) \neq f(\boldsymbol{x_j}) \) and  
\[
d_b(\boldsymbol{p_i},\boldsymbol{p_j}) < 2t - b + 2 - d_b(\boldsymbol{x_i},\boldsymbol{x_j}).
\]
Consequently, we obtain  
\[
d_b(\Phi(\boldsymbol{x_i}), \Phi(\boldsymbol{x_j})) \leq d_b(\boldsymbol{x_i}, \boldsymbol{x_j}) + d_b(\boldsymbol{p_i}, \boldsymbol{p_j}) + (b-1) < 2t + 1.
\]
This contradicts that $\Phi$ defines a function-correcting $b-$symbol code. Hence, 
\[
N_b(\boldsymbol{B}_f^{(1)}(t)) \leq r_b^f(k,t).
\]

Next, we establish the reverse inequality \( r_b^f(k,t) \leq N_b(\boldsymbol{B}_f^{(2)}(t)) \). Let $P = \{\boldsymbol{p_1}, \ldots, \boldsymbol{p_{q^k}}\}$ be a \(\boldsymbol{ B}_f^{(2)}(t) \) irregular $b-$symbol distance code of length \( N_b(\boldsymbol{B}_f^{(2)}(t)) \), and define the encoding function  
\[
\Phi: \mathbb{F}_q^k \to \mathbb{F}_q^{k+r}, \quad \boldsymbol{x_i} \mapsto (\boldsymbol{x_i}, \boldsymbol{p_i}).
\]  
For every pair \( i, j \in \{1, \ldots, q^k\} \) with \( f(\boldsymbol{x_i}) \neq f(\boldsymbol{x_j}) \), we have  
\[
d_b(\Phi(\boldsymbol{x_i}), \Phi(\boldsymbol{x_j})) \geq d_b(\boldsymbol{x_i}, \boldsymbol{x_j}) + d_b(\boldsymbol{p_i}, \boldsymbol{p_j}) - (b-1).
\]  
Since \( P \) is a \( \boldsymbol{ B}_f^{(2)}(t) \) irregular $b-$symbol distance code, it satisfies  
\[
d_b(\boldsymbol{p_i}, \boldsymbol{p_j}) \geq 2t + b - d_b(\boldsymbol{x_i}, \boldsymbol{x_j}),
\]  
which implies  
\[
d_b(\Phi(\boldsymbol{x_i}), \Phi(\boldsymbol{x_j})) \geq 2t + b - (b-1) = 2t + 1.
\]  
Thus, \( \Phi \) defines a function-correcting \( b \)-symbol code for \( f \) with redundancy \( r = N_b(\boldsymbol{B}_f^{(2)}(t)) \), yielding  
\[
r^f_b(k, t) \leq N_b(\boldsymbol{B}_f^{(2)}(t)).
\]  

This completes the proof.  
    \end{proof}


\subsection{Bounds on Optimal Redundancy for Generic Functions}

        In the previous section, lower and upper bounds were obtained for optimal redundancies of function-correcting $b$-symbol codes in terms of $N_b(\boldsymbol{B_f^{(1)}}(t))$ and $N_b(\boldsymbol{B_f^{(2)}}(t))$. These bounds can be simplified further by considering subsets of message vectors instead of the whole space ${\mathbb{F}^k_q}$.\\
First, we obtain a simplified lower bound on $r_b^f(k,t)$ in the following corollary of the theorem.\\
       \begin{corollary}:\label{corollary 3.3} For any arbitrary subset $ \{\boldsymbol{x}_1, \ldots, \boldsymbol{x}_M\} \subseteq {\mathbb{F}_q^k} $, optimal redundancy of a function-correcting $b$-symbol code for function $f$ is bounded below by $N_b(\boldsymbol{B_f^{(1)}}(t,\boldsymbol{x}_1, \ldots, \boldsymbol{x}_M))$ i.e.
        \[
            r^f_b(k, t) \geq N_b(\boldsymbol{B_f^{(1)}}(t,\boldsymbol{x}_1, \ldots, \boldsymbol{x}_M))
        \]
        and for function $f$ with $|Im(f)| \geq 2$ and $t > b-1$ \\
        \[
            r^f_b(k, t) \geq 2(t - b + 1)
        \]
        \end{corollary}
        \begin{proof} The first statement of the result is clear from the fact that every  $ \boldsymbol{B}_f^{(1)}(t)$ irregular $b-$symbol distance code is a $\boldsymbol{B_f^{(1)}}(t,\boldsymbol{x}_1, \ldots, \boldsymbol{x}_M)$ irregular $b-$symbol distance code. So,
        \[
            N_b(\boldsymbol{B_f^{(1)}}(t)) \geq N_b(\boldsymbol{B_f^{(1)}}(t,\boldsymbol{x}_1, \ldots, \boldsymbol{x}_M)).
        \]
Combining the above inequality with the before-established inequality $N_b(\boldsymbol{B}_f^{(1)}(t)) \leq r_b^f(k,t)$ , we get our required result i.e. 
         \[
            r^f_b(k, t) \geq N_b(\boldsymbol{B_f^{(1)}}(t,\boldsymbol{x}_1, \ldots, \boldsymbol{x}_M)).
         \]
Now consider the particular case where $|Im(f)| \geq 2$. Then there exist two distinct message vectors, say, $\boldsymbol{x}_i$ and $\boldsymbol{x}_j$ such that $f(\boldsymbol{x}_i) \neq f(\boldsymbol{x}_j)$ and $d_b(\boldsymbol{x}_i, \boldsymbol{x}_j) = b$. Let $t> b-1$ then,
        \[
            r^f_b(k, t) \geq N_b(\boldsymbol{B_f^{(1)}}(t,\boldsymbol{x}_i, \boldsymbol{x}_j)) = N_b(2,2t-b+2-b) = 2(t - b + 1).
        \]
        where the last equality can be obtained using a $2t - 2b + 2$ length repetition code \{(0, . . . , 0), (1, . . . , 1)\}.\\
        Now, we work our way through simplifying the upper bound on $r^f_b(k, t)$ but for that, we first introduce $b$-distance between function values which then leads us to define the function $b$-distance matrix.
        \end{proof}
        

        \begin{definition} Let \(f_1, f_2 \in Im(f)\), then the minimum $b-$symbol distance between two information vectors that evaluate to \( f_1 \) and \( f_2 \)  gives the $b-$symbol distance between the function values $f_1$ and $f_2$, i.e.,
        \[
        d_b^f (f_1, f_2) = \min_{x_1, x_2 \in \mathbb{F}_q^k} d_b (\boldsymbol{x}_1, \boldsymbol{x}_2) \quad \text{s.t.} \quad f(\boldsymbol{x}_1) = f_1 \text{ and } f(\boldsymbol{x}_2) = f_2.
        \]
        \end{definition}
        

        \begin{definition} The function $b$-distance matrices for function $f$ are square matrices of order $E  = |\text{Im}(f)|$ denoted by $\boldsymbol{B_f^{(1)}}(t,f_1,\ldots, f_E)$ and $\boldsymbol{B_f^{(2)}}(t,f_1,\ldots, f_E)$ respectively and whose entries are given by:
        \[
            [\boldsymbol{B_f^{(1)}}( t, f_1, \dots, f_E)]_{ij} = 
            \begin{cases} 
                [2t - b + 2 - d_b^f (f_i, f_j)]^{+}, & \text{if } i \neq j, \\ 
                0, & \text{otherwise}.
            \end{cases}
        \]
        and
        \[
            [\boldsymbol{B_f^{(2)}}( t, f_1, \dots, f_E)]_{ij} = 
            \begin{cases} 
                [2t + b  - d_b^f (f_i, f_j)]^{+}, & \text{if } i \neq j, \\ 
                0, & \text{otherwise}.
            \end{cases}
        \]
        \end{definition}
        With the help of the above definitions, we can now establish simplified upper bounds on optimal redundancy for any arbitrary function.
        
        \begin{theorem}
          The optimal redundancy for any arbitrary function $f$ defined in ${\mathbb{F}^k_q}$ is bounded above by $N_b(\boldsymbol{B_f^{(2)}}( t, f_1, \dots, f_E)).$
        \end{theorem}
        \begin{proof}
Let $\mathcal{P} = \{\boldsymbol{p_1}, \boldsymbol{p_2}, \ldots, \boldsymbol{p_E}\}$ be an $\boldsymbol{B_f^{(2)}}( t, f_1, \dots, f_E)-$irregular $b-$symbol distance code, i.e., for every $i$ and $j \in \{1, 2, \ldots, E \}$
\[
            d_b(\boldsymbol{p_i}, \boldsymbol{p_j}) \geq 2t + b - d_b^f (f_i, f_j).
 \]
Using the irregular \( b- \)symbol distance code defined above, we construct a function-correcting \( b \)-symbol code for the function \( f \) by assigning the same redundancy vector to all information vectors that produce the same functional value. Specifically, we assign \( \boldsymbol{p_i} \) to all inputs that satisfy \( f(\boldsymbol{x_i}) = f_i \).

To verify that this indeed forms a function-correcting \( b \)-symbol code, consider two arbitrary information vectors \( \boldsymbol{x_i}, \boldsymbol{x_j} \in \mathbb{F}_q^k \) such that \( f(\boldsymbol{x_i}) = f_i \) and \( f(\boldsymbol{x_j}) = f_j \), where \( f_i \neq f_j \). Then,
\begin{align*}
    d_b(\text{Enc}(\boldsymbol{x_i}), \text{Enc}(\boldsymbol{x_j})) & \geq d_b(\boldsymbol{x_i},\boldsymbol{x_j})+ d_b(\boldsymbol{p_i}, \boldsymbol{p_j}) - (b-1) \\
    & \geq d_b(\boldsymbol{x_i},\boldsymbol{x_j}) + 2t + b - d_b(\boldsymbol{x_i},\boldsymbol{x_j}) - b + 1 \\
    & = 2t + 1.
\end{align*}
Thus, by definition, the above construction yields a function-correcting \( b \)-symbol code.

Since such a code \( \mathcal{P} \) exists only if its length is at least \( N_b(\boldsymbol{B_f^{(2)}}(t, f_1, \dots, f_E)) \), we conclude that
\[
r_b^f(k,t) \leq N_b(\boldsymbol{B_f^{(2)}}(t, f_1, \dots, f_E)).
\]
        \end{proof}


\section{Bounds on $N_b(\textbf{B})$}

    As we observed, the optimal redundancy of function-correcting $b-$symbol distance codes is closely related to the length of irregular $b-$symbol distance code. Consequently, obtaining bounds on $N_b(\boldsymbol{B})$ also provides valuable insight to determine the optimal redundancy.

    Below we establish a lower bound on $N_b(\boldsymbol{B})$\\
    \begin{lemma}\label{Lemma 4.1}For any matrix $\boldsymbol{B} \in \mathbb{N}_0^{M \times M}$, we have
    \[
        N_b(\boldsymbol{B}) \ge
        \begin{cases}
        \frac{2q^b}{(q^b - 1)M^2} \sum_{i,j, \, i<j} [\boldsymbol{B}]_{ij}, & \text{if }  M \equiv 0 \pmod{q^b}, \\[10pt]
        \frac{2q^b}{(q^b - 1)(M^2 - 1)} \sum_{i,j, \, i<j} [\boldsymbol{B}]_{ij}, & \text{if } M \equiv 1 \pmod{q^b}, \\[10pt]
        \frac{2q^b}{M^2(q^b - 1) - m(q^b - m)} \sum_{i,j, \, i<j} [\boldsymbol{B}]_{ij}, & \text{if } M \equiv m \pmod{q^b},\text{ } m>1
        \end{cases}
    \]
    \end{lemma}
    \begin{proof} Let $\mathcal{P} = \{ \boldsymbol{p}_1, \ldots, \boldsymbol{p}_M\}$ be a $\boldsymbol{B}$-irregular $b-$symbol distance code of length $r$.\\
    Also, let $\boldsymbol{P}$ be a matrix of order $M \times r$ whose rows are the $b-$symbol read vectors of codewords of $\mathcal{P}$ and $\mathcal{S} = \sum_{i,j}d_b(\boldsymbol{p}_i, \boldsymbol{p}_j)$.\\
    Let $\alpha = (\alpha_1, \ldots, \alpha_b), \beta = (\beta_1, \ldots, \beta_b)$ and $M_l(\alpha)$ denote number of times $\alpha$ $\in \mathbb{F}_q^b$ appears in $l$th column of $\boldsymbol{P}$. Then,\\
    \begin{align*}
        \mathcal{S} & = \sum_{l=1}^r\sum_{\alpha \in \mathbb{F}_q^b }M_l(\alpha)[M- M_l(\alpha)] \\
                    & = \sum_{l=1}^r \left( M\sum_{\alpha \in \mathbb{F}_q^b }M_l(\alpha) - \sum_{\alpha \in \mathbb{F}_q^b }M_l^2(\alpha)\right)  \\
                    & = \sum_{l=1}^r \left( M^2 - \sum_{\alpha \in \mathbb{F}_q^b }M_l^2(\alpha)\right)\\
                    & = 2\sum_{l=1}^r \left( \sum_{\substack{\alpha, \beta \in \mathbb{F}_q^b   \\ \alpha \neq \beta}} M_l(\alpha)M_l(\beta) \right)
    \end{align*}
    
    so,
     \[
        \sum_{i,j, i \leq j}d_b(\boldsymbol{p}_i, \boldsymbol{p}_j) = \frac{1}{2}\mathcal{S} = \sum_{l = 1}^{r}\left( \sum_{\substack{\alpha, \beta \in \mathbb{F}_q^b \\ \alpha \neq \beta}}M_l(\alpha)M_l(\beta)\right).
     \] 
    \textbf{Case 1: }If $M \equiv 0 \pmod{q^b}$ then,\\
    \[ \sum_{l = 1}^{r}\left( \sum_{\substack{\alpha, \beta \in \mathbb{F}_q^b \\ \alpha \neq \beta}}M_l(\alpha)M_l(\beta)\right) \text{ is maximized when each of } M_l(\alpha) = \frac{M}{q^b}.\text{ Hence we have }\]\\
    \begin{align*}
        \sum_{i,j,i<j}[\boldsymbol{B}]_{ij} \leq \sum_{i, j, i<j}d_b(\boldsymbol{p}_i, \boldsymbol{p}_j) \leq \binom{q^b}{2}\left(\frac{M}{q^b}\right)^2 r = \frac{(q^b - 1)M^2}{2q^b}r \\
    \end{align*}
    We have\\
    \[
        N_b(D) \geq \frac{2q^b}{(q^b - 1)M^2}\sum_{i,j,i<j}[\boldsymbol{B}]_{ij}
    \]
    \textbf{Case 2: }If $M \equiv 1 \pmod{q^b}$ then,\\
    \[ \sum_{l = 1}^{r}\left( \sum_{\substack{\alpha, \beta \in \mathbb{F}_q^b \\ \alpha \neq \beta}}M_l(\alpha)M_l(\beta)\right) \text{ is maximized when } q^b - 1 \text{ of } M_l(\alpha) = \frac{M-1}{q^b} \] \[\text{ and any one of them is equal to } \frac{M-1}{q^b} + 1\] \\
    So, we have\\
    \[
        \sum_{i,j,i<j}[\boldsymbol{B}]_{ij} \leq \sum_{i, j, i<j}d_b(\boldsymbol{p}_i, \boldsymbol{p}_j) \leq \left[\binom{q^b - 1}{2}\left(\frac{M - 1}{q^b}\right)^2 +\left( q^b - 1\right) \left(\frac{M - 1}{q^b}\right)\left(\frac{M-1}{q^b}+1\right)\right]r  \\
    \]
    \[
        \implies \sum_{i,j,i<j}[\boldsymbol{B}]_{ij} \leq  \frac{(q^b - 1)(M^2 - 1)}{2q^b}r
    \]
    So, we have
    \[
        N_b(D) \geq \frac{2q^b}{(q^b - 1)(M^2 - 1)}\sum_{i,j,i<j}[\boldsymbol{B}]_{ij}
    \]
    \textbf{Case 3}: If $M \equiv m \pmod{q^b}$ where $m > 1$ then,\\
    \[ 
        \sum_{l = 1}^{r}\left( \sum_{\substack{\alpha, \beta \in \mathbb{F}_q^b\\ \alpha \neq \beta }}M_l(\alpha)M_l(\beta)\right) \text{ is maximized when } q^b - m \text{ of } M_l(\alpha) = \frac{M-1}{q^b} \] \[\text{ and rest $m$ of them is equal to } \frac{M-1}{q^b} + 1
    \] \\
    So, we have\\
    \begin{align*}
        \sum_{i,j,i<j}[\boldsymbol{B}]_{ij} & \leq \sum_{i, j, i<j}d_b(\boldsymbol{p}_i, \boldsymbol{p}_j) \\
                                            & \leq \left[\binom{q^b - m}{2}\left(\frac{M - m}{q^b}\right)^2  +  m\left( q^b - m\right) \left(\frac{M - m}{q^b}\right)\left(\frac{M-m}{q^b}+1\right)  +  \binom{m}{2}\left(\frac{M-m}{q^b}+1\right)^2 \right]r  \\
                                            & = \left[ \frac{M^2(q^b - 1) - m(q^b - m)}{2q^b} \right]r\\
    \end{align*}
    So, we have
       \[
             N_b(\boldsymbol{B}) \geq \frac{2q^b}{M^2(q^b - 1) - m(q^b - m)}\sum_{i,j,i<j}[\boldsymbol{B}]_{ij}
       \]
    \end{proof}
    
     \begin{remark} The Lemma \ref{Lemma 4.1} generalizes the Plotkin-like bound\cite{yang2016construction}. Take $[\boldsymbol{B}]_{ij} = D$ for $i \neq j$ then from Lemma \ref{Lemma 4.1} we get, $N_b(M,D) \geq \frac{q^b(M-1)}{M(q^b - 1)}$, which is clearly a variant of the Plotkin-like bound for $b$-symbol codes.
     \end{remark}
     

    \begin{lemma}\label{Lemma 4.3} For any distance matrix \( \boldsymbol{B} \in \mathbb{N}^{M\times M}_0 \) and any permutation \( \pi : \{1, 2, \ldots, M\} \rightarrow \{1, 2, \ldots, M\} \), we have
    \[
    N_b(\boldsymbol{B}) \leq \min_{r \in \mathbb{N}} \left\{ r \ \middle|\ q^r > \max_{j \in \{1, 2, \ldots, M\}} \sum_{i=1}^{j-1} |\mathcal{B}^{b}_{r}([\boldsymbol{B}]_{\sigma(i)\sigma(j)} - 1) |\right\},
    \]
    where \( \mathcal{B}^{b}_r(t) = \{ y \in \mathbb{F}_q^r\ | d_b(\boldsymbol{x},\boldsymbol{y}) \leq t \} \).
    \end{lemma}
    \begin{proof}We present an iterative way of selecting valid codewords for constructing an irregular $b-$symbol distance code for any distance matrix $\textbf{B}$($\boldsymbol{B_b-}$code).\\
    At first, for the sake of simplicity, choose $\pi$ as the identity permutation. Select any arbitrary vector $\boldsymbol{p_1} \in \mathbb{F}_q^r$ as the first codeword of the $\boldsymbol{B_b-}$ code. Make a selection for the second codeword $\boldsymbol{p_2}$ so that the distance requirement condition, $d_b(\boldsymbol{p_1},\boldsymbol{p_2}) \geq [\boldsymbol{B}]_{12}$ is satisfied. The existence of $\boldsymbol{p_2}$ is guaranteed if $q^r > | \mathcal{B}^{b}_{r}([\boldsymbol{B}]_{12} - 1)| $. Then choose the third codeword $\boldsymbol{p_3}$ s.t. $d_b(\boldsymbol{p_1}, \boldsymbol{p_3}) \geq [\boldsymbol{B}]_{13}$ and $d_b(\boldsymbol{p_2}, \boldsymbol{p_3}) \geq [\boldsymbol{B}]_{23}$. For $\boldsymbol{p_3}$, the existence is guaranteed if $q^r > | \mathcal{B}^{b}_{r}([\boldsymbol{B}]_{13} - 1)| + |\mathcal{B}^{b}_{r}([\boldsymbol{B}]_{23} - 1)| $. Continuing this process, we can select all codewords up to $\boldsymbol{p_M}$ under the given condition of the lemma. Also, since codewords can be selected in any order, the lemma holds for any permutation $\pi$.
    \end{proof}
    \begin{remark} The Lemma \ref{Lemma 4.3} is a generalization of the Gilbert-Varshamov bound \cite{song2018sphere} and for $[\boldsymbol{B}]_{ij} = D$, the lower bound in Lemma 4.3 becomes  
    \[
    N_b(M,D) \leq \min_{r \in \mathbb{N}} \left\{ r \ \middle|\ q^r > (M-1) |\mathcal{B}^{b}_{r}([\boldsymbol{B}]_{ij})| \right\},
    \]
     which is clearly a variant of the Gilbert-Varshamov bound.
     \end{remark}

     The next lemma connects $N_{b+1}(M,D)$ with $N_b(M,D)$.\\
     \begin{lemma}
          Let \( M \) and \( D \) be positive integers with \( M > q^{b} \) and \( D \geq b-1 \). Then
    \[
        N_{b+1}(M, D) \leq N_b(M, D - 1) .
    \]
     \end{lemma}
     \begin{proof}
         Let \( r = N_b(M, D - 1) \). Then there exists a code \( \mathcal{P} = \{\boldsymbol{ p_1}, \dots, \boldsymbol{p_M} \} \) of length \( r \) such that
    \[ 
        d_b(\boldsymbol{p_i}, \boldsymbol{p_j}) \geq D -  1 \hspace{2mm} \forall  i, j \in \{1, 2, \ldots, M\}  \text{ with }  i \neq j .
    \]  
    By the Singleton bound for the $b-$symbol metric \cite{ding2018maximum}, we have \\
    \[
        q^{r - (D - 1) + b} \geq M > q^{b}
    \] which implies 
    \[
        r - (D -  1) + b> b 
        \implies r > D - 1.
    \]
    If \( d_b(\boldsymbol{p_i}, \boldsymbol{p_j}) <  r  \), then
    \[
    d_{b+1}(\boldsymbol{p_i}, \boldsymbol{p_j}) \geq d_b(\boldsymbol{p_i}, \boldsymbol{p_j}) +  1 \geq D - 1 + 1 = D.
    \]
    If \( d_b(\boldsymbol{p_i}, \boldsymbol{p_j}) = r\), then \( d_{b+1}(\boldsymbol{p_i}, \boldsymbol{p_j}) = d_b(\boldsymbol{p_i}, \boldsymbol{p_j}) =  r \geq D \). \\
    It follows that \( N_{b+1}(M, D) \leq r = N_b(M, D -  1) \).
    \end{proof}
    The next lemma gives a relation between the shortest length of irregular distance codes $N_H(M,D)$ and the shortest length of irregular $b-$symbol distance code  $N_b(M,D)$.   
  
    \begin{lemma}\label{Lemma 4.5} Let \( M \) and \( D \) be positive integers with \( M > q^{b-1} \) and \( D \geq b \). Then
    \[
        N_b(M, D) \leq N_H(M, D - b + 1) .
    \]
    \end{lemma}
    \begin{proof} Let \( r = N_H(M, D - b + 1) \). Then there exists a code \( \mathcal{P} = \{\boldsymbol{ p_1}, \dots, \boldsymbol{p_M} \} \) of length \( r \) such that
    \[ 
        d_H(\boldsymbol{p_i}, \boldsymbol{p_j}) \geq D - b + 1 \hspace{2mm} \forall  i, j \in \{1, 2, \ldots, M\}  \text{ with }  i \neq j .
    \]  
    By the Singleton bound in the Hamming metric \cite{huffman2010fundamentals}, we have \\
    \[
        q^{r - (D - b + 1) + 1} \geq M > q^{b-1}
    \] which implies 
    \[
        r - (D - b + 1) + 1 > b - 1
        \implies r > D - 1.
    \]
    If \( d_H(\boldsymbol{p_i}, \boldsymbol{p_j}) \leq r - b + 1 \), then
    \[
    d_b(\boldsymbol{p_i}, \boldsymbol{p_j}) \geq d_H(\boldsymbol{p_i}, \boldsymbol{p_j}) + b - 1 \geq D - b + 1 + b -1 = D.
    \]
    If \( d_H(\boldsymbol{p_i}, \boldsymbol{p_j}) > r - b +1 \), then \( d_b(p_i, p_j) = r \geq D \). \\
    It follows that \( N_b(M, D) \leq r = N_H(M, D - b + 1) \).\\\\
    To utilize the above relation to obtain bounds on $N_b(M, D)$ we first establish upper bounds on $N_H(M, D)$ for q-ary codes.
    \end{proof}
    
    \begin{lemma}(\cite[Def.4.32]{Horadam+2007})\label{Lemma 4.6} Let $D \in \mathbb{N}$ and $q \geq 2$ such that $q-1$ divides $D$. Let H be a generalized Hadamard matrix of order $\frac{qD}{q-1}$ i.e. $ H = GH \left( q, \frac{D}{q-1} \right)$ and let M be a positive integer satisfying $M \leq \frac{q^2D}{q-1}$ then,
    \[
        N_H(M,D) \leq \frac{qD}{q-1}.
    \]
    \end{lemma}
    Combining the result of Lemma \ref{Lemma 4.5} and Lemma \ref{Lemma 4.6}, we get the following result.
    
    \begin{lemma}\label{Lemma 4.7} Let $D \geq b$  and $q \geq 2$ be positive integers such that $ q-1 \text{ divides } D - b + 1$. Then, for a generalized Hadamard matrix of order $\frac{q(D-b+1)}{q-1}$ and $M \leq \frac{q^2(D-b+1)}{q-1}$ we have
    \[
        N_b(M,D) \leq \frac{q(D-b+1)}{q-1}.
    \]
    \end{lemma}
    
    \begin{lemma}\label{Lemma 4.8} Let $M,D \in \mathbb{N}$ and $q \geq 2$ be such that $ D \geq 10 $ and  $M \leq D^2$ then,
    \[
         N_H(M,D) \leq \frac{q(D-1)}{q(1 - \sqrt{\frac{\ln D}{D}})-1}.
    \]
    \end{lemma}
   \begin{proof} 
    From \cite[Lemma 4.7.2]{ash2012information}, an upper bound can be determined for the size of a Hamming ball \( V_q(r, D-1) = \sum_{i=0}^{D-1} \binom{r}{i}(q-1)^i \), provided \( D - 1 \leq \frac{(q-1)r}{q} \). Specifically, this bound is given by:  
    \[ V_q(r, D-1) \leq q^r e^{-\frac{q^2r}{2(q-1)}\left( \frac{q-1}{q} - \frac{D-1}{r} \right)^2}. 
    \]  
      Furthermore, using the Gilbert-Varshamov (GV) bound established in Remark 4.4, we know that \( q^r > M V_q(r, D-1) \). Combining these two inequalities yields:  
    \[ q^r > M q^r e^{-\frac{q^2r}{2(q-1)}\left( \frac{q-1}{q} - \frac{D-1}{r} \right)^2}. \]  
    This inequality implies the existence of a code of size \( M \), minimum distance \( D \), and length \( r \).  
    
    Now, setting \( D = r \frac{q-1}{q} - \epsilon r + 1 \) for some \( 0 \leq \epsilon < \frac{q-1}{q} \), we derive a \([M, D]\) code of length \( r \) that satisfies \( M \leq e^{\frac{q^2r}{2(q-1)}\epsilon^2} \). Choose \( \epsilon = \sqrt{\frac{\ln r}{r}} \), which leads to:  
    \[ r = \frac{q(D-1)}{q\left(1 - \sqrt{\frac{\ln r}{r}}\right) - 1}. \]  
    For \( r \geq 10 \), we have \( \epsilon \leq \frac{1}{2} \leq \frac{q-1}{q} \). Furthermore, since \( \frac{\ln D}{D} \geq \frac{\ln r}{r} \) for \( r \geq D \geq 3 \), the inequality in the lemma holds for \( D \geq 10 \).
    \end{proof}
    From Lemmas \ref{Lemma 4.8} and \ref{Lemma 4.5}, we can conclude the following lemma.
    
    \begin{lemma}\label{Lemma 4.9} 
    Let $q \geq 2 $ and  $M,D \in \mathbb{N}$ such that $ D \geq 9 + b $ and  $M \leq (D-b+1)^2$ then,
    \[
         N_b(M,D) \leq \frac{q(D-b)}{q\left(1 - \sqrt{\frac{\ln D - b + 1}{D - b + 1}}\right)-1}.
    \]
    \end{lemma}
    In the paper \cite{xia2024function} a relation between $N_p(M,D)$ and $N_p(\boldsymbol{D})$ was presented where $N_p(\boldsymbol{D})$ denotes the smallest possible length of a $\boldsymbol{D}-$irregular-pair distance codes for a $M$ order matrix $\boldsymbol{D}$ with nonnegative integer entries and $N_p(M,D)$ is the special case of $N_p(\boldsymbol{D})$ where all entries of $\boldsymbol{D}$ are equal to D. We also present a similar relation for $\boldsymbol{B}_b-$codes.
    
    \begin{lemma}
    Let $\boldsymbol{B}$ be a matrix of $M$ order with nonnegative integer entries, and let $D_{max} = max_{i,j}[\boldsymbol{B}]_{ij}$. If $D_{max} \leq D,$, then $N_b(\boldsymbol{B}) \leq N_b(M,D).$ 
    \end{lemma}   
    The relation in the above lemma gives us a way to give an upper bound on $ N_b(\boldsymbol{B})$ using Lemmas \ref{Lemma 4.7}  and \ref{Lemma 4.9}.
    
    \begin{corollary}\label{corollary 4.11}
    Let $\boldsymbol{B}$ be a matrix of order $M$ with non-negative integer entries, and let $D_{max} = max_{i,j}[\boldsymbol{B}]_{ij}$ be such that $D_{max} \leq D$. If $D \geq b$ and $q \geq 2$ are positive integers such that $q - 1$ divides $D - b + 1$ and a generalized Hadamard matrix of order $\frac{q(D-b+1)}{q-1}$ and $M \leq \frac{q^2(D-b+1)}{q-1}$, then $N_b(\boldsymbol{D}) \leq  \frac{q(D-b+1)}{q-1}.$
    \end{corollary}
    
    \begin{corollary}\label{corollary 4.12}  Let $\boldsymbol{B}$ be a matrix of order $M$ with non-negative integer entries, and let $D_{max} = max_{i,j}[\boldsymbol{B}]_{ij}$ be such that $D_{max} \leq D$. If $ D \geq 9 + b $ and $M \leq (D-b+1)^2$ then for $q \geq 2$,
    \[
         N_b(\boldsymbol{B}) \leq \frac{q(D-b)}{q\left(1 - \sqrt{\frac{\ln D - b + 1}{D - b + 1}}\right)-1}.
    \]
    \end{corollary}


\section{$b$-Symbol Locally Binary Functions}
    In \cite{lenz2023function} and \cite{xia2024function}, the authors introduced a family of locally binary functions with respect to the Hamming distance and the symbol-pair distance respectively. Building on this concept, we introduce the $b$-symbol locally binary functions for $b$-symbol distance.
    
    \begin{definition} A function ball of a function $f$ with $b$-symbol radius $\rho$ around $\boldsymbol{x} \in {\mathbb{F}^k_q}$ is defined as
    \begin{equation}
        B^f_b(\boldsymbol{x},\rho) = \{ f(\boldsymbol{\Bar{x}}): \boldsymbol{\Bar{x}} \in  {\mathbb{F}^k_q} \text{ and } d_b(\boldsymbol{x}, \boldsymbol{\Bar{x}}) \leq \rho \}
    \end{equation}
    \end{definition}

    \begin{definition} A function $f: {\mathbb{F}^k_q} \rightarrow Im(f)$ is called a $\rho$-$b$-locally binary function, if for all $\boldsymbol{x} \in {\mathbb{F}^k_q}$,
        \[
          |B^f_b(\boldsymbol{x},\rho)| \leq 2
        \]
    \end{definition}
    
    \begin{lemma} For any $2t-b-$locally binary function such that $t > b-1$ we have $2(t- b + 1) \leq r_b^f(k,t) \leq 2t - b + 1$.
    \end{lemma}
    \begin{proof}The lower bound directly results from Corollary \ref{corollary 3.3} We provide the construction of a function-correcting $b-$symbol code for $2t$-$b$-locally binary function whose length of redundancy is $2t-b+1$ and thus establish the stated upper bound on the optimal redundancy.\\For $\boldsymbol{x} \in \mathbb{F}^k_q$, let for some non-zero symbol $a$ in $\mathbb{F}_q$,
    \[
        p(x) = 
        \begin{cases}
            a, & if \quad f(\boldsymbol{x}) = max\left( B_b^f(\boldsymbol{x}, 2t)\right)\\
            0, & otherwise
        \end{cases}
    \]
    and 
    \[
        Enc(\boldsymbol{x}) = (\boldsymbol{x}, (p(\boldsymbol{x})^{2t-b+1}) )
    \]
    where $(p(\boldsymbol{x})^{2t-b+1})$  means $(2t - b + 1)$-fold repetition of the bit $p(\boldsymbol{x})$.\\
    We show that the above encoding function defines a function-correcting $b-$symbol code.\\
    Let $\boldsymbol{x}$ and $\boldsymbol{x'} \in \mathbb{F}^k_q$ such that $f(\boldsymbol{x}) \neq f(\boldsymbol{x'})$.\\
    Case 1: $d_b(\boldsymbol{x},\boldsymbol{x}') > 2t$, \\
    from the construction we can easily see $d_b(Enc(\boldsymbol{x}), Enc(\boldsymbol{x'})) \geq d_b(\boldsymbol{x},\boldsymbol{x'}) \geq 2t + 1$ \\
    So, when $d_b(\boldsymbol{x},\boldsymbol{x'}) > 2t$, $d_b(Enc(\boldsymbol{x}), Enc(\boldsymbol{x'})) \geq 2t + 1$ \\
    Case 2: $d_b(\boldsymbol{x},\boldsymbol{x'}) \leq 2t$, then $p(\boldsymbol{x}) \neq p(\boldsymbol{x'})$. \\
    If  $d_H(Enc(\boldsymbol{x}), Enc(\boldsymbol{x'})) > k-b+1$ then   $d_b(Enc(\boldsymbol{x}), Enc(\boldsymbol{x'})) = k + 2t - b +1 \geq 2t+1.$ 
    Otherwise,
    \begin{align*}
     d_b(Enc(\boldsymbol{x}), Enc(\boldsymbol{x'})) & \geq d_H(Enc(\boldsymbol{x}), Enc(\boldsymbol{x'})) + (b - 1), \\ 
                         & \geq  2t- b + 2 + b - 1\\
                         & = 2t + 1 
    \end{align*}
    Thus, the above encoding function is indeed a function correcting $b$-symbol code for $2t-b-$symbol locally binary function and hence $r_b^f(k,t) \leq 2t - b + 1$.
    \end{proof}    


\section{$b$-Symbol Weight Functions}

\begin{figure}[t]
        \centering
        \includegraphics[width=0.8\linewidth]{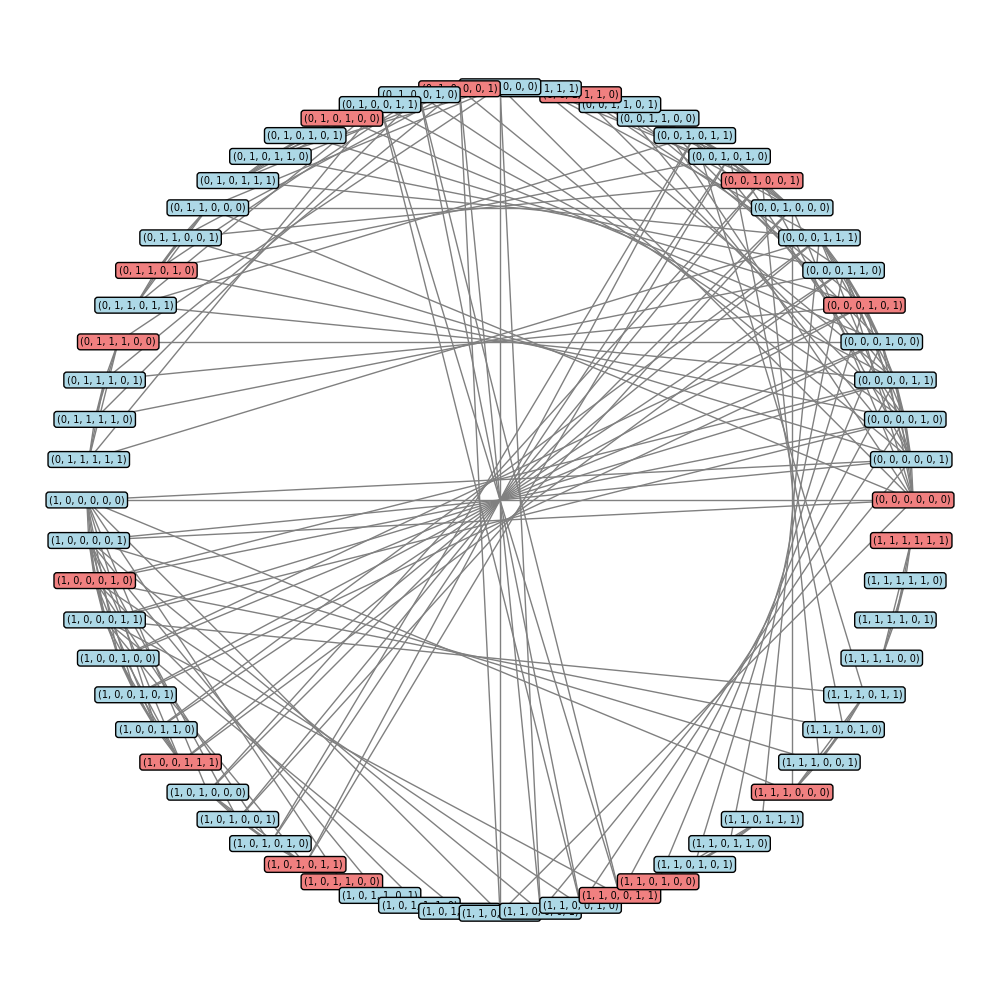}
        \caption{ Graph $G_{wt_3}^3(k,t,r)$ for $k = 4, t = 2, r =2$. The independent set of size 16 is highlighted in red boxes, forming the function-correcting $b-$symbol codes for $wt_3$ function. }
        \label{fig:3}
    \end{figure}

    In this section, we go through a class of functions called \textbf{$b$-weight functions} and try to establish upper and lower bounds on the optimal redundancy of function-correcting $b$-symbol code designed for these functions using the previously established results.\\
    Let $wt_b : \mathbb{F}^k_q \rightarrow \{ 0, b, b+1, \ldots, k\}$ be the $b-$symbol function defined in $\mathbb{F}^k_q$. This implies $E = |Im(wt_b)| = k - b + 2$.\\
    Using the graphical approach described in Section 3.1, the function-correcting $b-$symbol codes can be obtained in the form of independent sets of graph $G_{wt_b}^b(k,r,t)$.  For example, Figure \ref{fig:3} shows the graph $G_{wt_3}^b(k,r,t)$ for $k = 4, t = 2, r = 2, b = 3$ over the binary field.

    We now present an upper bound and lower bound for the optimal redundancy $r_b^{wt_b}(k,t)$ of the function-correcting $b$-symbol codes for the function $wt_b$ in the following two lemmas using Theorem \ref{Theorem 3.2} and Corollary \ref{corollary 3.3}, respectively.
    
    \begin{lemma}\label{Lemma 6.1}
        Let $g(\boldsymbol{x}) = wt_b(\boldsymbol{x})$, $g_1 = 0$ and $ g_i = b + (i - 2) \quad \forall \quad 2 \leq i \leq k-b+2 $.\\
        Let us consider the function $b-$symbol distance matrix $B_{g}^{(2)}(t,g_1, g_2, \dots, g_{k-b+2})$ of $g(\boldsymbol{x})$ then the optimal redundancy of function-correcting $b$-symbol code for $wt_b$ is bounded above by $N_b(B_{g}^{(2)}(t, g_1, \ldots, g_{k-b+2}))$ i.e.
        \[
            r_b^{g}(k,t) \leq N_b(B_{g}^{(2)}(t, g_1, \ldots, g_{k-b+2})) 
        \]
     where
        \[
            [B_{wt_b}^{(2)}(t, g_1, \ldots, g_{k-b+2})]_{ij} = 
            \begin{cases}
                0,                  & i=j\\
                2t - max\{i,j\} + 2,  & i \neq j \text{ and }  i = 1 \text{ or } j = 1 \\
                2t,                 &  i, j  > 1 \text{ and } 0 < |i - j| < b   \\   
                [2t + b - |i-j|]^+, & i, j > 1 \text{ and } |i - j| \geq b
            \end{cases}
        \]
    \end{lemma}
    \begin{proof}
        Let, $d_b^g(g_i, g_j) = min_{x_1, x_2 \in \mathbb{F}^k_q}d_b(\boldsymbol{x}_1, \boldsymbol{x}_2)$
            \[
                 \text{ where, }\\
                 g_i = wt_b(\boldsymbol{x_1}) = 
                 \begin{cases}
                     b + i - 2, & i \geq 2\\
                     0,          & i = 1.
                 \end{cases}  \text{ and } g_j = wt_b(\boldsymbol{x}_2) = 
                 \begin{cases}
                     b + j - 2, & j \geq 2\\
                     0,          & j = 1.
                 \end{cases}
            \] 
        We know, for $i \neq j$\\
        
            \[
                \text{ if } i = 1 \quad or \quad j = 1, \text{ then } d_b^g(g_i, g_j) = b + max\{i,j\} - 2.
            \]
            
            \[
                \text{ if } i,j > 1 \quad and \quad |i-j| \leq b-1 \text{ then } d_b^g(g_i, g_j) = b \text{ and }
            \]
            
            \[
                \text{ if } i,j >1 \quad and \quad |i-j| \geq b \text{ then } d_b^g(g_i, g_j) = |i-j|
            \]
        So, the function $b-$symbol distance matrix for $wt_b$ is :\\
            \[
                [B_{wt_b}^{(2)}(t, g_1, \ldots, g_E)]_{ij} = 
                \begin{cases}
                    0,                  & i=j\\
                    2t - max\{i,j\} + 2,  & i \neq j \text{ and }  i = 1 \text{ or } j = 1 \\
                    2t,                 &  i, j  > 1 \text{ and } 0 < |i - j| < b   \\   
                    [2t + b - |i-j|]^+, & i, j > 1 \text{ and } |i - j| \geq b
                \end{cases}
            \]
    \end{proof}
        
    \begin{lemma}\label{Lemma 6.2}
        Let $\boldsymbol{x_i} = (a^i0^{k-i})$ where $a$ is a nonzero symbol over the field $\mathbb{F}_q$ and $i \in \{0, 1, \ldots, k-b+1\}$. Then, optimal redundancy $r_b^{wt_b}(k,t)$ of $b-$symbol weight function is bounded below by $N_b(B_{wt_b}^{(1)}(t, \boldsymbol{x_0}, \ldots, \boldsymbol{x_{k-b+1}}))$ i.e.
        
            \[
                r_b^{wt_b}(k,t) \geq N_b(B_{wt_b}^{(1)}(t, \boldsymbol{x_0}, \ldots, \boldsymbol{x_{k-b+1}}))
            \]
            
        where \\
        
            \[
                [B_{wt_b}^{(1)}(t, \boldsymbol{x_0}, \ldots, \boldsymbol{x_{k-b+1}})]_{ij} = 
                \begin{cases}
                    0, & if \quad i = j,\\
                    [2t - 2b + 3 - |i-j|]^+, & if \quad i \neq j.
                \end{cases}
            \]
        \end{lemma}
    \begin{proof} 
        We know from the Corollary \ref{corollary 3.3} that $r_b^{wt_b}(k,t) \geq N_b(B_{wt_b}^{(1)}(t, \boldsymbol{x_0}, \ldots, \boldsymbol{x_{k-b+1}}))$.\\Now, the entries of $b-$symbol distance matrix $B_{wt_b}^{(1)}(t, \boldsymbol{x_0}, \ldots, \boldsymbol{x_{k-b+1}})$ is given by
            \[
                [B_{wt_b}^{(1)}(t, \boldsymbol{x_0}, \ldots, \boldsymbol{x_{k-b+1}})]_{ij} = 
                \begin{cases}
                    [2t - b + 2 - d_b(\boldsymbol{x_i}, \boldsymbol{x_j})]^+, & \quad wt_b(\boldsymbol{x_i}) \neq wt_b(\boldsymbol{x_j}),\\
                    0, & otherwise.
                \end{cases}
            \]
        As $d_b(\boldsymbol{x_i}, \boldsymbol{x_j}) = |i - j| + b - 1$ and $wt_b(\boldsymbol{x_i}) \neq wt_b(\boldsymbol{x_j}) \text{ for } i \neq j$ so the above $b-$symbol distance matrix simplifies to
        \[
          [B_{wt_b}^{(1)}(t, \boldsymbol{x_0}, \ldots, \boldsymbol{x_{k-b+1}})]_{ij} = 
            \begin{cases}
                0, & if \quad i = j,\\
                [2t - 2b + 3 - |i-j|]^+, & if \quad i \neq j.
            \end{cases}  
        \]
    \end{proof}
    We present an example for the function $b-$symbol distance matrix $B_{wt_b}^{(2)}(t, g_1, \ldots, g_E)$ and $b-$symbol distance matrix $B_{wt_b}^{(1)}(t, \boldsymbol{x_0}, \ldots, \boldsymbol{x_{k-b+1}})$ for particular values of k, b and t.
    \begin{example} 
        For $q = 2, k = 6, t = 3 \text{ and } b = 3$, the entries of the $B_{wt_3}^{(2)}(3, g_1, \ldots, g_5)$ of order 5 can be calculated using the definition in Lemma \ref{Lemma 6.1}:

        \begin{align*}
             &1. [B_{wt_3}^{(2)}(3, g_1, \ldots, g_E) = 0 ]_{ij} = 0, \quad if \quad i = j,\\
             &2. [B_{wt_3}^{(2)}(3, g_1, \ldots, g_E) = 0 ]_{ij} = 2t - max \{i, j\} + 2 ,\quad if \quad i \neq j  \text{ and }i = 1 \text{ or } j = 1 ,\\
             &3. [B_{wt_b}^{(2)}(3, g_1, \ldots, g_E) = 0 ]_{ij}= 2t ,\quad if \quad i, j > 1 \quad and \quad 0 < |i - j| < b ,\\
             &4. [B_{wt_b}^{(2)}(3, g_1, \ldots, g_E) = 0 ]_{ij} = [2t + b - |i - j|]^+ ,\quad if \quad i, j > 1  \quad and \quad  |i - j| \geq b .\\
        \end{align*}

        Hence, the matrix \( B_{wt_b}^{(2)}(t, g_1, \ldots, g_5) \) for $q = 2,$ \( k = 6 \), \( b = 3 \), and \( t = 3 \) is:

        \[ B_{wt_b}^{(2)}(t, g_1, \ldots, g_5) = 
            \begin{bmatrix}
            g(\boldsymbol{x}) & \vline & 0 & 3 & 4 & 5 & 6 \\ \hline 
            0 & \vline & 0 & 6 & 5 & 4 & 3  \\
            3 & \vline & 6 & 0 & 6 & 6 & 6  \\
            4 & \vline & 5 & 6 & 0 & 6 & 6 \\
            5 & \vline & 4 & 6 & 6 & 0 & 6 \\
            6 & \vline & 3 & 6 & 6 & 6 & 0 \\
            \end{bmatrix}
        \]
        Similarly from Lemma  \ref{Lemma 6.2} the entries of $B_{wt_3}^{(1)}(3, \boldsymbol{x_0}, \ldots, \boldsymbol{x_{4}})$ are as follow:
            \[ B_{wt_3}^{(1)}(3, \boldsymbol{x_0}, \ldots, \boldsymbol{x_{4}}) =
                \begin{bmatrix}
                    0 & 2 & 1 & 0 & 1\\
                    2 & 0 & 2 & 1 & 0\\
                    1 & 2 & 0 & 2 & 1 \\
                    0 & 1 & 2 & 0 & 2 \\
                    1 & 0 & 1 & 2 & 0\\
                \end{bmatrix}   
            \]
    \end{example}
    We now present another lower bound on $r_b^{wt_b}(k,t)$ as a function of $t$ using the already established lower bound on $N_b(\boldsymbol{B})$ in Lemma \ref{Lemma 4.1}.

    \begin{corollary} When $k>t$ , we have
        \[
            r_b^{wt_b}(k,t) \geq \frac{5q^b(t - b+2)(t-b+1)}{ 3(q^b - 1)(t -b+3)}.
        \]
    \end{corollary}
    \begin{proof}
        Let $\boldsymbol{x_i} = (a^i0^{k-i})$ where $a$ is a non-zero symbol over the field $\mathbb{F}_q$ and $i \in \{0, 1, \ldots, k-b+1\}$ and $\mathcal{P} = \{ \boldsymbol{p_0}, \boldsymbol{p_1}, \ldots, \boldsymbol{p_{k-b+1}} \}$ be a $B_{wt_b}^{(1)}(t, \boldsymbol{x_0}, \ldots, \boldsymbol{x_{k-b+1}})$-code of length $N_b(B_{wt_b}^{(1)}(t, \boldsymbol{x_0}, \ldots, \boldsymbol{x_{k-b+1}}))$.\\
        We consider the principal sub-matrix $\boldsymbol{B'}$ obtained by selecting first $t-b+3$ rows and corresponding $t-b+3$ columns from matrix $B_{wt_b}^{(1)}(t, \boldsymbol{x_0}, \ldots, \boldsymbol{x_{k-b+1}})$\\
        Let P be a $(t-b+3) \times (k-b+2)$ matrix given by 
        \[
            P = 
            \begin{bmatrix}
                I_{t-b+3} & 0
            \end{bmatrix}
        \]
        where $I_{t-b+3}$ is an identity matrix of order $t-b+3$ and $0$ is a null matrix.
        Also, let Q be a $(k-b+2) \times (t-b+3)$ matrix given by 
        \[
            Q = 
            \begin{bmatrix}
                I_{t-b+3} \\
                0
            \end{bmatrix}
        \]
        Then $\boldsymbol{B'}$ is given by 
        \[
           \boldsymbol{B'} = PB_{wt_b}^{(1)}(t, \boldsymbol{x_0}, \ldots, \boldsymbol{x_{k-b+1}})Q.
        \]
        We can see that the first $t-b+1$ vectors of $\mathcal{P}$ are a $\boldsymbol{B'}_b-code$. By Lemma \ref{Lemma 4.1}
        \begin{align*}
            N_b(\boldsymbol{B'}) & \geq \frac{2q^b}{(q^b - 1)(t-b+3)^2} \sum^{t-b+3}_{i=1} \sum^{t-b+3}_{j=i+1} [\boldsymbol{B'}]_{ij}\\ 
                                 &  =   \frac{2q^b}{(q^b - 1)(t-b+3)^2} \sum^{t-b+3}_{i=1} \sum^{t-b+3}_{j=i+1} [B_{wt_b}^{(1)}(t, \boldsymbol{x_0}, \ldots, \boldsymbol{x_{k-b+1}})]_{ij}\\
                                 &  =   \frac{2q^b}{(q^b - 1)(t-b+3)^2} \sum^{t-b+2}_{l=1} (2t - 2b + 3 - l)(t-b+3 - l)\\
                                 &  =   \frac{5q^b(t - b+2)(t-b+1)}{ 3(q^b - 1)(t -b+3)}                         
        \end{align*}
        Combining the above inequality with inequality in Lemma \ref{Lemma 6.2} we get
        \[
            r_b^{wt_b}(k,t) \geq  \frac{5q^b(t - b+2)(t-b+1)}{ 3(q^b - 1)(t -b+3)}.
        \]
    \end{proof}    
    
    \begin{corollary} For $t \geq \lceil \frac{9+b}{2} \rceil$ and $ b \leq k \leq (2t - b + 1)^2,$ we have
        \[
           r_b^{wt_b}(k,t)  \leq \frac{q(2t-b)}{q\left(1 - \sqrt{\frac{\ln (2t - b + 1)}{2t - b + 1}}\right)-1}.
        \]
    \end{corollary}
    \begin{proof} 
        In Lemma \ref{Lemma 6.1}, we see that the maximum value $[B_{wt_b}^{(2)}(t, g_1, \ldots, g_E)]_{ij}$ can assume is $2t$ and since $t \geq \lceil \frac{9+b}{2} \rceil$ and $k \leq (2t - b + 1)^2$ so by corollary \ref{corollary 4.12} we have,
        \[
            N_b(B_{wt_b}^{(2)}(t, g_1, \ldots, g_E)) \leq \frac{q(2t-b)}{q\left(1 - \sqrt{\frac{\ln 2t - b + 1}{2t - b + 1}}\right)-1}.
        \]
        Therefore, $r_b^{wt_b}(k,t) \leq N_b(B_{wt_b}^{(2)}(t, g_1, \ldots, g_E)) \leq \frac{q(2t-b)}{q\left(1 - \sqrt{\frac{\ln 2t - b + 1}{2t - b + 1}}\right)-1}$.
    \end{proof}
    By Corollary \ref{corollary 3.3} we know that for any generic function $f$ with $|Im f| \geq 2$, the optimal redundancy of function-correcting $b-$symbol codes is bounded below by $2(t-b+2)$. The next lemma presents an improved lower bound on optimal redundancy in the case of $b-$symbol weight functions.

    \begin{lemma}
        For the $b-$symbol weight function, the optimal redundancy of the function-correcting $b-$symbol code is bounded below by $2t - b + 1$.
        \[
            r_b^{wt_b}(k,t) \geq 2t - b + 1.
        \]
    \end{lemma}
    \begin{proof}
        Consider a function-correcting $b-$symbol code for a $b-$symbol weight function whose redundancy $r = r_b^{wt_b}(k,t) < 2t - b + 1$ and the encoding function are given by
        \[
            Enc: \mathbb{F}_q^k \rightarrow \mathbb{F}_q^{k+r}, \boldsymbol{x} \mapsto (\boldsymbol{x},p(\boldsymbol{x}))
        \]
        Let $\boldsymbol{x} = (0, 0, \ldots, 0)$ and $\boldsymbol{x'} = (0, 0, \ldots, 1).$ Then 
        \[
            d_b(Enc(\boldsymbol{x}),Enc(\boldsymbol{x'})) \leq d_b(\boldsymbol{x},\boldsymbol{x'}) + d_b(p(\boldsymbol{x}),p(\boldsymbol{x'}))  \leq b + 2t - b = 2t.
        \]
        which is a contradiction. Hence, $r_b^{wt_b}(k,t) \geq 2t - b + 1$.
    \end{proof}
    

    Next, we provide an explicit method to construct the function-correcting $b-$symbol code for the $b-$symbol weight function.
    
    \begin{construction}
        We define the encoding function, $Enc(\boldsymbol{x}) = (\boldsymbol{x}, \boldsymbol{p}_{wt_{b}(\boldsymbol{x})+1})$ for all $\boldsymbol{x} \in \mathbb{F}_q^k$ with the help of smod operator($ c \text{ smod }d  = ((c - 1) \text{ mod } d ) + 1$)  where redundancy vectors $\boldsymbol{p's}$ are defined as follows: Let $\boldsymbol{p}_1, \ldots, \boldsymbol{p}_{2t+1}$ be a collection of  $2t+1$ codewords with minimum distance $2t$ i.e.
        \[
            min\{d_b(\boldsymbol{p_i},\boldsymbol{p}_j) :  i,j \leq 2t + 1, i \neq j\} = 2t.
        \]
        The rest of the redundancy vectors are defined with the help of the smod operator $\boldsymbol{p_i} = \boldsymbol{p}_{i smod (2t+1)}$.\\
        We claim that the above encoding function defines a function-correcting $b-$symbol code for $b-$symbol weight functions.
    \end{construction}
    \begin{proof}
        For any $\boldsymbol{x}, \boldsymbol{x}' \in \mathbb{F}_q^k$ such that $wt_b(\boldsymbol{x}) \neq wt_b(\boldsymbol{x'})$,
        \[
            d_b(Enc(\boldsymbol{x}), Enc(\boldsymbol{x'})) \geq d_b(\boldsymbol{x},\boldsymbol{x'}) + d_b(p(\boldsymbol{x}),p(\boldsymbol{x'})) - (b - 1) \geq b + 2t - b + 1 = 2t + 1.
        \] 
        Therefore, the above encoding function defines a function-correcting $b-$symbol code for $b-$symbol weight functions.
    \end{proof}


\section{ $b$-Symbol Weight Distribution Functions}

    $b-$symbol distribution functions are an important class of functions in coding theory, as these functions provide valuable insights into the \textbf{error detection and correction} capabilities of a code.\\ 
    For a parameter of choice $T \in \mathbb{N}$ such that $T \geq \lfloor \frac{b+2}{2} \rfloor$, let $f(\boldsymbol{x})$  be a $b$-weight distribution function  defined over $\mathbb{F}_q^k$ as follows 
    \[
        f(\boldsymbol{x}) = \Delta_b^T(\boldsymbol{x}) = \lfloor \frac{wt_b(\boldsymbol{x})}{T} \rfloor.
    \]
For mathematical simplicity, we consider only those $T$ for which $T$ divides $k +1$. In this way $f(\boldsymbol{x})$ behaves as a step function based on  $b-$symbol weight, with the steps equal to $E = \frac{k+1}{T}$.\\
    We present a construction of function-correcting \( b-\)symbol codes for the \( b \)-symbol weight distribution function. This construction is based on assigning the same redundancy vector to information vectors that share the same \( b \)-symbol weight with respect to \( \text{smod } T \).
    
     
    \begin{construction}
        Let $Enc_{f}$ define an encoding map from $\mathbb{F}_q^k$ to $\mathbb{F}_q^{k+2t}$ as follows \[
            Enc_{f}(\boldsymbol{x}) = (\boldsymbol{x}, \boldsymbol{p}_{(wt_b(\boldsymbol{x})+1) smod T})
        \]
where $\boldsymbol{p}_{(wt_b(\boldsymbol{x})+1) smod T} \in \mathbb{F}_q^{2t}$ and for some nonzero element $a \in \mathbb{F}_q$ it is defined as follows:
        \[
            p_i = 
            \begin{cases}
                (a^{i-1}0^{2t-i+1}), & if \quad 1 \leq i \leq 2t+1,\\
                (a^{2t}), & if \quad 2t+2 \leq i \leq T .
            \end{cases}
        \]
    \end{construction}
    
    \begin{lemma} 
        For any given $k,T$ with $k \geq b$ and $T \geq \lfloor \frac{b+2}{2} \rfloor$ such that $T$ divides $k+1$ and $T$ greater than or equal to $2t + 1$, $Enc_{f}$ defines a function-correcting $b-$symbol code for the $b$-weight distribution function with redundancy equal to 2t.
    \end{lemma}
    \begin{proof}
 For any \( \boldsymbol{x_1} \) and \( \boldsymbol{x_2} \) such that \( f(\boldsymbol{x_1}) \neq f(\boldsymbol{x_2}) \), we consider the following two cases:\\ 
 Case 1: \( d_b(\boldsymbol{x_1}, \boldsymbol{x_2}) \geq 2t + 1 \) . \\
 In this case, the result is straightforward, as \( d_b(Enc_{f}(\boldsymbol{x_1}), Enc_{f}(\boldsymbol{x_2})) \geq 2t + 1 \) holds directly.  \\
Case 2: \( d_b(\boldsymbol{x_1}, \boldsymbol{x_2}) \leq 2t \)  \\
We note that \( T \geq 2t + 1 \). So, $ d_b(\boldsymbol{x_1}, \boldsymbol{x_2}) < T $  i.e., $ wt_b(\boldsymbol{x_1} - \boldsymbol{x_2}) < T \implies |\frac{wt_b(\boldsymbol{x_1})}{T} - \frac{wt_b(\boldsymbol{x_2})}{T}| < 1.$ 
Since $f(\boldsymbol{x_1}) \neq f(\boldsymbol{x_2})$ this implies that \( f(\boldsymbol{x_1}) \) and \( f(\boldsymbol{x_2}) \) differ by exactly one. Without loss of generality, assume that \( f(\boldsymbol{x_1}) = m-1 \) and \( f(\boldsymbol{x_2}) = m \).  
        
Hence, the \( b \)-weights of \( \boldsymbol{x_1} \) and \( \boldsymbol{x_2} \) can then be expressed as:  
        \[
        wt_b(\boldsymbol{x_1}) = (m - 1)T + w_1, \quad wt_b(\boldsymbol{x_2}) = mT + w_2.
        \]
        Subcase 2.1: \( T > b \)  \\
         For \( m = 1 \):  \( w_1 \in \{0, b, \ldots, T-1\} \)  and  \( w_2 \in \{0, 1, \ldots, T-1\} \)  \\
         For \( m \geq 2 \):  \( w_1, w_2 \in \{0, 1, \ldots, T-1\} \)  \\
         Subcase 2.2: \( T < b \) and \( T \geq \lfloor \frac{b+2}{2} \rfloor \)  \\
         For \( m = 1 \):  \( w_1 = 0 \)  and \( w_2 \in \{0, b, \ldots, T-1\} \)  \\
         For \( m = 2 \):  \( w_1 \in \{0, b, \ldots, T-1\} \)  and  \( w_2 \in \{0, 1, \ldots, T-1\} \) .
         First,  we try to prove for $T = 2t + 1,$ 
         In this particular case, $\boldsymbol{p}(\boldsymbol{x_1}) =  \boldsymbol{p}_{w_1 + 1} = (a^{w_1}0^{2t-w_1})$ and $\boldsymbol{p}(\boldsymbol{x_2}) =  \boldsymbol{p}_{w_2 + 1} = (a^{w_2}0^{2t-w_2})$.\\
         So,  
         \begin{align*}
         d_b(Enc_{f}(\boldsymbol{x_1}), Enc_{f}(\boldsymbol{x_2}))  & \geq d_b(\boldsymbol{x_1}, \boldsymbol{x_2}) + d_b(\boldsymbol{p}(\boldsymbol{x_1}), \boldsymbol{p}(\boldsymbol{x_2})) - (b-1).\\
         & \geq|wt_b(\boldsymbol{x_1}) - wt_b(\boldsymbol{x_2})| + |w_1 - w_2| + b - 1 -( b - 1) \\
         & \geq (T - w_1 + w_2) +( w_1 - w_2 ).\\
         & = T = 2t + 1.\\
         \end{align*}
           For $T > 2t + 1,$ we consider the following three cases:\\
        \begin{itemize}
               \item if $w_1 \geq 2t$, then $d_b(p(\boldsymbol{x_1}), p(\boldsymbol{x_2})) = 2t - w_2 + b -1$. So,
               \begin{align*}
                   d_b(Enc_{f}(\boldsymbol{x_1}), Enc_{f}(\boldsymbol{x_2}))  & \geq T - w_1 + w_2 + 2t - w_2 + b - 1 - (b - 1)\\ 
                   & \geq T + 2t - w_1.\\
                   & \geq T > 2t + 1.
               \end{align*} 
               \item if $w_2 \geq 2t$, then $d_b(p(\boldsymbol{x_1}), p(\boldsymbol{x_2})) = 2t - w_1 + b - 1$.
               \item $w_1 < 2t$ and $w_2 < 2t$, then $d_b(p(\boldsymbol{x_1}), p(\boldsymbol{x_2})) = |w_1 - w_2|$. These cases can be proven similarly.
        \end{itemize}
    \end{proof}
    From the above construction and Corollary \ref{corollary 3.3} we conclude the following.
         
    \begin{corollary}
        For any given $k,t,T$ with $k \geq b$, $t \geq 1$, and $T \geq \lfloor \frac{b+2}{2} \rfloor$ such that $T$ divides $k+1$ and $T$ greater than equal to $2t + 1$,  we have $2(t - b +1) \leq r_b^{\Delta_b^t}(k,t) \leq 2t.$
    \end{corollary}

\section{Redundancy Comparison with Classical $b-$Symbol Codes } 

    \begin{table}\label{Table_1}
         \centering
         \resizebox{\textwidth}{!}{\begin{tabular}{|c|c|c|}
         \hline
            Functions & Classical $3$-symbol Code & function-correcting $3-$symbol Codes \\[11pt]
        \hline
             $2t-b-$locally binary functions & $\lfloor \frac{t}{3} \rfloor log_q(k + 2t - 2)$ & $2t - 2$ \\[11pt]
        \hline
             $b-$symbol weight functions & $\lfloor \frac{t}{3} \rfloor log_q(k + 2t - 2)$ &  $\frac{q(2t-3)}{q\left( 1 - \sqrt{\frac{\ln(2t-2)}{2t - 2}}\right) - 1}$ \\[11pt]
        \hline
             $$b-$$symbol weight distribution functions & $\lfloor \frac{t}{3} \rfloor log_q(k + 2t - 2)$ & $2t$  \\[11pt]
        \hline
         \end{tabular}}
         \caption{A Comparison of Redundancy in Classical 3-Symbol Codes and Function-Correcting 3-Symbol Codes}
    \end{table} 

    \begin{table}
         \centering
         \resizebox{\textwidth}{!}{\begin{tabular}{|c|c|c|c|c|c|}
         \hline
            Functions & \makecell{Redundancy\\FCC(b = 1)} & Functions & \makecell{Redundancy\\FCSPC(b = 2)} & Functions & \makecell{Redundancy\\Function-correcting\\ 3-symbol codes(b = 3)} \\[11pt]
        \hline
             \makecell{$2t-$locally binary\\ functions} & $r_f(k, t) = 2t $ & \makecell{$2t-$pair$-$locally \\ binary functions} & $2t-2 \leq r_p^f(k,t) \leq 2t-1$ & \makecell{$2t-3-$locally binary\\ functions} &  $2t-4 \leq r_b^f(k,t) \leq 2t-2$\\[11pt]
        \hline
             \makecell{Hamming weight\\ functions} & \makecell{$\frac{10t}{3} - \frac{10}{3} \leq r_wt(k,t)$ \\ $\leq \frac{4t}{1 - 2\sqrt{\ln(2t)/2t}}$} & Pair weight functions & \makecell{ $\frac{20t^3 -20t}{9(t+1)^2} \leq r_p^{w_p}(k,t)$ \\ $\leq \frac{4t-4}{1 - 2\sqrt{\ln(2t-1)/2t-1}}$} & $b-$symbol weight functions & \makecell{$\frac{40t^2-120t+80}{21t} 
            \leq r_3^{wt_b}(k,t)$ \\ $\leq \frac{4t - 6}{1 - 2\sqrt{\frac{\ln (2t - 2)}{2t-2}}}$}  \\[13pt]
        \hline
             \makecell{Hamming weight \\distribution functions} & $r_\Delta(k,t) = 2t$ & \makecell{Pair weight\\ distribution functions} & $2t - 2 \leq r_p^{\delta_p^T}(k, t) \leq 2t$ & \makecell{$b-$symbol weight distribution\\ function} &   $2(t - 2) \leq r_b^{\Delta_b^t}(k,t) \leq 2t$ \\[11pt]
        \hline
         \end{tabular}}
         \caption{Redundancy Comparison between FCC, FCSPC and Function-Correcting $3-$Symbol Codes}
     \end{table} 

    To demonstrate that using function-correcting \( b \)-symbol codes offers a distinct advantage over classical $b-$symbol codes in terms of reduced redundancy for certain classes of functions, we compare the redundancy of the two approaches across all the aforementioned function classes.\\
    We know $q^n \geq q^k|\mathcal{B}^b_n(t)|$, from the sphere packing bound for $b-$symbols, mentioned in Lemma \ref{sphere_packing}. From the asymptotic analysis of  $|\mathcal{B}^3_n(t)|$ in  Proposition \ref{b-symbol_ball} we conclude that $|\mathcal{B}^3_t(n)| = \Theta(n^{\lfloor \frac{t}{3}  \rfloor })$. Combining the above results, we get
    \[
        n-k \geq \lfloor \frac{t}{3}  \rfloor \cdot log_q n.
    \]
    Also, we know from the singleton bound for the $b-$symbol codes 
    \[
            n \geq k  + 2t + 1 - b.
    \]
    Thus, we have
    \[
            n-k \geq \lfloor \frac{t}{3}  \rfloor \cdot log_q ( k  + 2t -2).
    \]
    Based on the previously derived lower bound for the redundancy of classical $b-$symbol codes, we provide the redundancy values for the classical 3-symbol in column 2 of Table 1, for each of the three classes of functions discussed earlier. Column 3 in the table lists the corresponding upper bounds on the redundancy for function-correcting $b-$symbol codes for these functions. For sufficiently large values of \( k \), it can be shown that reduced redundancy can be achieved using function-correcting \( b \)-symbol codes. For instance, if we fix \( q = 2 \) and take \( k > 2^{2b} - 2t - 1 + b \), then \( \lfloor \frac{t}{b} \rfloor \log_2(k + 2t + 1 - b) > 2t - 1 \). Therefore, by comparing columns 2 and 3, it is evident that for the aforementioned classes of functions, function-correcting \( b \)-symbol codes achieve reduced redundancy.
    

\section{Conclusion}
    Function-correcting b-symbol codes represent a more advanced and efficient error correction strategy, aligning well with the requirements of modern data storage systems that rely on $b-$symbol read channels. They offer practical advantages in terms of speed, capacity, and error recovery, which are crucial to meeting the demands of today's high-performance storage devices.
    We developed function-correcting \( b \)-symbol codes for functions over finite fields to ensure the protection of function values evaluated on messages against errors introduced by a \( b- \)symbol read channel. By establishing bounds on the length of irregular-distance codes and irregular \( b- \)symbol distance codes over finite fields, and leveraging the relationship between function-correcting \( b- \)symbol codes and irregular \( b- \)symbol distance codes, we derived both lower and upper bounds on the optimal redundancy of generic functions. For certain classes of functions, we have demonstrated that these codes can achieve significantly lower redundancy compared to classical \( b- \)symbol codes.


\appendix
\section{Appendix}
\label{appendix}
\begin{figure}
    \centering
    \includegraphics[width=0.5\linewidth]{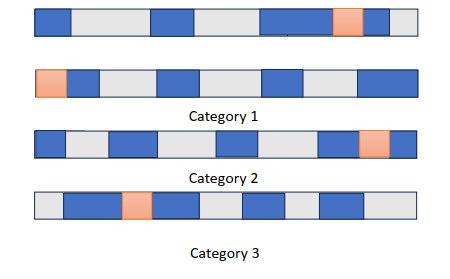}
    \caption{The figure illustrates the layouts of $(k,l, L_j)$ subsets for $L_j = 4$ and $j = 1$, where the lighter coloured rectangles represent unoccupied boxes, darker rectangles represent occupied rectangles and orange represents size 1 unoccupied subset. Category 1 depicts layouts where neither the occupied nor the unoccupied rectangles wrap around the ends. Category 2 represents cases where an occupied rectangle wraps around the ends and Category 3 shows the case when an unoccupied rectangle wraps around. }
    \label{fig:4}
\end{figure}
In this appendix, we give a proof of the expression $D_j(k,l, L_j)$ used in Proposition \ref{b-symbol_ball} using one of the elementary counting results, which states that the number of ways to divide $n$ identical elements into $k$ numbered bins that are non-empty is $\binom{n-1}{k-1}$\cite{vanLintWilson2001} .\\
Categorize the subsets that meet the $(k,l, L_j)$ specification into three categories as shown in figure \ref{fig:4}. For any $L_j$, the dark-coloured rectangles represent elements in the size-$l$ subset. The light-coloured rectangles along with the orange-coloured rectanle represent elements that are not in the subset. In addition, the orange rectangles are the size-$1$ rectangles.\\
\textbf{Category 1:} This category considers the case where neither dark-coloured rectangles nor light-coloured rectangles wrap around the ends. Of the $L_j$ rectangles that are not in the subset, the j rectangles have exactly $1$ element, and the number of ways to choose these j rectangles is $\binom{L_j}{j}$. The remaining $L_j - j$ rectangles should contain at least $2$ elements, so the number of ways to divide the remaining $k-l-j$ elements into $L_j - j$ rectangles is $\binom{k-l-L_j-1}{L_j - j - 1}$. For dark-coloured rectangles, the $l$ elements can be divided among $L_j$ rectangles so that no rectangle is empty in $\binom{l - 1}{L_j - 1}$ ways.\\
\textbf{Category 2: }In this category consider dark-coloured rectangles wrapped around the ends. In this case, there are $L_j + 1$ rectangles in which $l$ elements can be divided into. Therefore the number of ways to do so is $\binom{l-1}{L_j}$. For light-coloured rectangles, the conditions remain the same as Category 1 and hence we have $\binom{k-l-L_j-1}{L_j - j - 1}\binom{L_j}{j}$.\\
\textbf{Category 3: }Consider the case where light-coloured rectangles can wrap around the ends. So, for the $k-l$ elements not in the subset, there are $L_j + 1$ rectangles to choose from. But the size-$1$ subsets have only $L_j$ choices as they cannot wrap around the ends. Hence, the number of ways to select the $j$ size-$1$ subsets is equal to $\binom{L_j}{j}$ and the remaining $k-l-j$ elements can be divided among $L_j - j + 1$ subsets in $\binom{k-l-L_j-1}{L_j - j}$ ways. As for the size-$l$ subsets, they can be divided into $L_j$ rectangles in $\binom{l-1}{L_j - 1}$ ways.\\
As the elements in the dark-coloured, light-coloured and orange-coloured rectangles can be independently grouped, the number of $(k,l, L_j)$ subsets from each category is the product of the number of light-coloured, dark-coloured and orange coloured groupings. Hence, we get $D_j(k,l, L_j) = 2\binom{k-l-L_j-1}{L_j - j - 1}\binom{L_j}{j}\binom{l-1}{L_j - 1} + \binom{k-l-L_j-1}{L_j - j - 1}\binom{L_j}{j}\binom{l-1}{L_j} + \binom{k-l-L_j-1}{L_j - j }\binom{L_j}{j}\binom{l-1}{L_j - 1} = \frac{k}{L_j}\binom{l-1}{L_j-1}\binom{k-l-L_j-1}{L_j-j-1}\binom{L_j}{j}. $



\bibliographystyle{unsrt} 
\bibliography{doc/main}


\end{document}